\documentclass[journal,letterpaper,10pt]{IEEEtran}

\usepackage{header}
\usepackage{enumitem}
\usepackage{url}
\usepackage{xurl}
\usepackage{xurl}
\usepackage{hyperref}
\hypersetup{colorlinks,allcolors=black}

\begin{document}

\title{Stochastic High-Resolution Parameter-Based Channel Modeling for Sub-Terahertz Outdoor Device-to-Device Propagation at 145 GHz}

\author{Zihang Cheng,~\IEEEmembership{Student Member,~IEEE},
        Naveed A. Abbasi,~\IEEEmembership{Member,~IEEE},
        Jorge Gomez-Ponce,~\IEEEmembership{Senior Member,~IEEE},
        Jianzhong (Charlie) Zhang,~\IEEEmembership{Fellow,~IEEE} and
        Andreas F. Molisch,~\IEEEmembership{Fellow,~IEEE}\vspace{-2em}
\thanks{ZC, NAA, and AFM are Dpt. of ECE, Univ. Southern Cal., Los Angeles, CA, \{zihangch,nabbasi,molisch\}@usc.edu. JGP is with ESPOL, Guayaquil, Ecuador. JCZ is with Samsung Research America, Plano, TX. Source code: \url{https://github.com/jolugome1989/subTHz-D2D-HRPE-simulator}}}

\maketitle

\begin{abstract} Since device-to-device (D2D) systems operating in the sub-THz regime are of interest for a variety of 6G applications, modeling their propagation channels is therefore essential. This paper presents high-resolution parameter estimation (HRPE) of specular multipath components (MPCs) as well as diffuse multipath components (DMCs) from an extensive measurement campaign. The extracted MPCs and DMCs are then clustered, and statistical models for both intra- and inter-cluster statistics are established. Among other insights, we find a large number (typically 10-50) of clusters; both the powers and densities of MPCs within each cluster show significant variations. DMC occurs both when associated with discrete MPCs and standalone. The paper provides a complete model that can generate double-directional impulse responses, and a MATLAB program implementing it is made available. Extensive verifications of ``derived'' (not directly modeled) parameters are provided as well. 
    
\end{abstract}


\vspace{-12pt}
\section{Introduction }

\Ac{D2D} communication is an important deployment scenario in 5G and 6G wireless systems, as the absence of infrastructure offers flexibility and low communication latency \cite{ansari20175g}. This is relevant for applications such as (i) public safety networks \cite{ali2016architecture}, (ii) military communications \cite{bajracharya20235g}, and (iii) \ac{V2V} communications \cite{zeadally2020tutorial}. In all of these applications, the special properties of (sub-)THz links, in particular the available high bandwidth and ability to perform precise beamforming even with low-form-factor antenna arrays, are particularly beneficial \cite{tataria20216g}.

Accurate characterization of the propagation channels based on, or verified by, measurements is the prerequisite for any system development, as the channel determines both fundamental performance limits and the effectiveness of particular transmission schemes \cite{molisch2022wireless}. While \ac{D2D} channels in lower frequency bands have been characterized in the past \cite{alobaidy2022wireless}, there is much less work in the sub-THz regime (see Sec. I.A). To remedy this, the current authors have conducted an extensive measurement campaign and derived from it models for pathloss, delay, and angular spread \cite{abbasi2022thz}, and a cluster-based channel model using Fourier evaluation \cite{cheng2025stochastic}.

While Fourier evaluation of measurements is fast and robust, and provides valuable insights, its resolution is fundamentally limited by the aperture (bandwidth and beamwidth) of the measurement equipment.\footnote{Improvement of resolution by using, e.g., antennas with narrower beamwidth is limited by the concomitant increase in measurement time when rotating-horn or lateral-displacement virtual array channel sounders are used, as almost always done in this frequency range \cite{han2022terahertz}.} Furthermore, interpolating channel behavior for beamwidths that are non-integer multiples of the measurement beamwidth is difficult. These fundamental problems can be eliminated or at least mitigated by \ac{HRPE}. In this approach, the channel is characterized by a parametric model, and the model parameters are determined, e.g., by \ac{ML} estimation, from the measurement results. Resolution in this case is limited by the \ac{CRLB} (which decreases with increasing \ac{SNR}) rather than the aperture, and the characteristics for any beamwidth or array configuration can be synthesized easily from the resulting channel model. The finer resolution and extraction of individual \acp{MPC} is important for a variety of applications, from verifying assumptions in existing channel models about the power distribution of \acp{MPC} to precise localization in \ac{NLoS} scenarios.

When creating channel models from \ac{HRPE} results, it is important to consider the {\em cluster structure} of the \acp{MPC}, i.e., the \acp{MPC} naturally form groups with similar parameters in the delay/angle space; this allows simpler modeling, as cluster-level statistics require fewer parameters than a model in which the statistics of each \ac{MPC} (and their correlations) need to be described. Clustering is thus widely used in both statistical and geometrical models \cite[Chapter 7]{molisch2022wireless}.
Furthermore, we need to account for the fact that the channel response includes both discrete \acp{MPC} representing plane waves, and \acp{DMC} \cite{jiang2022survey}. The latter can carry a significant fraction of the total received energy\textemdash in particular at higher frequencies, where many surfaces are rough in comparison to the wavelength. Ignoring the \ac{DMC} complicates the model (if its contributions are approximated by a large number of low-power discrete \acp{MPC}) and/or leads to erroneous system analysis (if its contributions are ignored in the model). Despite its importance, \ac{DMC} is extracted and modeled only in a small fraction of the channel modeling literature.  

\vspace{-10pt}
\subsection{Related Work}

{\em Clustered channel models} constitute a fundamental framework for characterizing multipath propagation in wireless communication systems. Clusters are groups of \acp{MPC} that have similar parameters (delays, \acp{DoA} and \acp{DoD}) while differing significantly in at least one parameter from other clusters. Clustered models are popular in academic papers \cite{saleh1987statistical,spencer2002modeling,gustafson2013mm,rappaport2017small} and standardization \cite{3GPPTR38.901,molisch2006cost259}. The seminal Saleh-Valenzuela (S-V) model describes {\em separately} the density of MPCs (Poisson) and their power decay (exponential) in the delay domain. However, for models that also incorporate the angular domain, including the generalization of the S-V model \cite{spencer2002modeling} and the 3GPP model \cite{3GPPTR38.901}, typically only {\em either} the MPC density {\em or} the power decay is modeled, while the other quantity is assumed constant. 


{\em Measurement evaluations combining \ac{HRPE} with diffuse scattering representations}: Evaluations of measurements using \ac{HRPE}, i.e., extracting specular \acp{MPC} with high temporal and angular resolution, are widely used. However, better accuracy can be achieved by adding a parametric \ac{DMC} model to the \acp{MPC} in the extraction process. Early work in this field modeled the \ac{DMC} as a single exponentially decaying cluster, either spatially uniform \cite{richter2005estimation,li2019modeling} or characterized by a \emph{von Mises} angular distribution \cite{kaske2009modelling}. Subsequent investigations revealed a more structured diffuse clustering behavior \cite{ribeiro2007joint}, and spatio-temporal extraction at 3.6 GHz demonstrated correlations between diffuse and specular components \cite{quitin2010diffuse}, suggesting that \ac{DMC} and \ac{MPC} clusters are inherently associated. To address this, we recently proposed an associated multi-cluster \ac{DMC} estimation algorithm \cite{cheng2024multi}. 

{\em Outdoor D2D Sub-THz channels} have been investigated in several papers. Ref. \cite{abbasi2020double} conducted the first outdoor sub-THz double-directional measurements, employing a \ac{D2D} configuration at 145 GHz, followed by a more extensive campaign \cite{abbasi2022thz} that also underlies the current paper. Subsequently,   \cite{undi2021angle} reported outdoor urban \ac{D2D} measurements at 300 GHz, which in \cite{keusgen2022sub} were compared to measurements at 158 GHz. Ref. \cite{selimis2024path} conducted measurements at 145 GHz in city center, residential, and suburban areas. Evaluations of these measurements were done only in terms of pathloss, delay spread, and angular spread. Only our recent \cite{cheng2025stochastic} provides a cluster-based model, which is based on Fourier evaluation. 
Besides these campaigns, other sub-THz investigations deal with microcellular setups and indoor environments, respectively. Even for those other environments,  
there are only a few \ac{HRPE} studies, and those remain predominantly focused on specular \ac{MPC} extraction. For example, \cite{li2024dss} applied the SAGE algorithm to real measurements at 300 GHz in an indoor corridor scenario, without incorporating an explicit diffuse scattering model. Consequently, to the best of our knowledge, the (sub-) THz literature currently lacks (i) a comprehensive HRPE-based channel modeling framework that jointly characterizes specular and diffuse components
, and (ii) a complete parameterized model for \ac{D2D} outdoor scenarios. 

\vspace{-10pt}
\subsection{Our Contributions}

To remedy these gaps in the literature, the current paper:

\begin{itemize}[leftmargin=*]
    \item Develops a framework for the \ac{HRPE} evaluation of sub-THz D2D channel measurements, and applies it to our extensive measurement campaign. Based on the extracted parameters, we propose a cluster-based stochastic channel model incorporating diffuse scattering. 
    \item Provides new insights into the statistical behavior of the channels, including (i) the angle-delay cluster structure of the DMC is partly associated with MPC clusters and partly standalone clusters, (ii) both the density and the strength of the \acp{MPC} depend on their location within the cluster, and the \ac{ADPS} is influenced by both, and (iii) the number of \acp{MPC} in a cluster is a random variable, and the powers of the \acp{MPC} within a cluster show significant variations; both of these results are contrary to the assumptions in the widely used 3GPP channel model. 
    \item Develops, and publicly releases, a D2D sub-THz channel simulator that provides results for (almost) arbitrary array and antenna configurations. 
\end{itemize}

\vspace{-12pt}
\section{Sub-THz D2D Measurement Campaign}
\vspace{-4pt}

To ease the interpretation of our parameter extraction results, this section provides a brief summary of our sounder setup and its calibration, and of the measurement environment and location. For more details, please see \cite{gomez2022directionally,abbasi2022thz}.

The measurements underlying our work were conducted with a custom channel sounder based on a \ac{VNA} with \ac{RoF} extensions (which allows measurements over long distances) and horn antennas mounted on precision rotors. For each horn orientation at the \ac{Tx} and \ac{Rx}, the VNA measures the channel transfer function by stepping through a 1 GHz measurement bandwidth in 1 MHz subcarrier spacing; this guarantees unambiguous range identification (no wrap-around) of the MPCs for excess runlength up to 300m, which was sufficient in the measured environments. The horn antennas have $13^{\circ}$ \ac{HPBW} in both azimuth and elevation and are rotated in azimuth only, since in D2D scenarios (Tx and Rx at the same height, namely 1.5m) propagation is dominantly in the horizontal plane (see \cite{cheng2025stochastic} for more details); the rotation step width is $10^{\circ}$, which guarantees overlap of the antenna patterns at adjacent rotation angles, which is essential for our \ac{HRPE} estimator. 

The measurements were conducted in an urban environment, i.e., the University Park Campus of the University of Southern California (USC). The specific environments are (i) an L-shaped park-like area (quad) surrounded by buildings, and containing vegetation as well as lampposts, concrete pillars, etc., and (ii) a street intersection with buildings on three of the corners, and a sports field (with metal fences) on the fourth. A total of 21 \ac{LoS} and 17 \ac{NLoS} location pairs were measured, with Tx - Rx in the range of 1--100m. More details and maps/photos of the environment and measurement locations can be found in \cite{abbasi2022thz}.

\vspace{-8pt}
\section{Signal Model and Parameter Extraction }
\vspace{-2pt}
\subsection{Parametric Channel Model}
\vspace{-2pt}
The transfer function of a double-directional channel \cite{steinbauer01} can be expressed as the superposition of plane waves (also known as \acp{MPC}) and \ac{DMC} contributions. We group the \acp{MPC} into $N_{\rm C}$ clusters, with the $c$-th cluster containing $M_c$ \acp{MPC}. 

The measured \ac{MIMO} transfer function matrix of the channel, $\mathbf{H}_{\rm mea}$, transmitted from $N_{\rm T}$ antennas and received by $N_{\rm R}$ antennas at a specific frequency point, $f_k$, can then be expressed as\footnote{We assume (i) static channels, (ii) single (vertical) polarization, (iii) Tx and Rx in the far field of the scatterers and of each other).}
\vspace{-16pt}
\begin{align}\label{eq:channel_model}     
    \mathbf{H}_{\rm mea}(f_k)=& \sum_{c=1}^{N_{\rm C}} \sum_{p=1}^{M_c} \gamma_{c,p} \tilde{\mathbf{g}}_{\rm R}(\bm{\Omega}_{c,p}^{\rm R}) \tilde{\mathbf{g}}_{\rm T}^{T}(\bm{\Omega}_{c,p}^{\rm T}) e^{-j 2 \pi f_k \tau_{c,p}} \notag \\
    &+ \mathbf{N}_{\rm DMC}(f_k) + \mathbf{N}(f_k),
\end{align}
    \vspace{-10pt}
    
    \noindent where $\gamma$, $\bm{\Omega}^{{\rm R}}$, $\bm{\Omega}^{{\rm T}}$ and $\tau$ are complex amplitude, \ac{DoA}, \ac{DoD} and propagation delay, respectively, of the $p$-th \ac{MPC} in the $c$-th cluster. The calibrated array patterns, $\tilde{\mathbf{g}}_{\rm R}(\bm{\Omega}_{{\rm R}}) \in \mathbb{C}^{N_{\rm R} \times 1}$ and $\tilde{\mathbf{g}}_{\rm T}(\bm{\Omega}_{{\rm T}}) \in \mathbb{C}^{N_{\rm T} \times 1}$, express the angular response from a certain direction. The additive noise, $\mathbf{N}(f_k) \in \mathbb{C}^{N_{\rm R} \times N_{\rm T}}$, is assumed to follow a zero-mean, circularly symmetric complex Gaussian distribution. $\mathbf{N}_{\rm DMC}(f_k)$ stands for the total contribution from the diffuse scattering, see below.

The clustering of the \acp{MPC} arises from the geometrical and physical structure of the propagation. The description of the \acp{MPC} is commonly split into their {\em inter-cluster} characteristics, which describe the relationships between the cluster centroids \cite{saleh1987statistical,3GPPTR38.901,molisch2003geometry,zhu2013cost}) channel, and the {\em intra-cluster} characteristics, where the intra-cluster {\em shape} in the delay-angle domain is assumed to be identical across clusters, while their {\em scaling} (which can be described by only a few scalar parameters) may differ from cluster to cluster.


\Ac{DMC} arises from diffuse scattering by rough surfaces, distributed scatterers such as foliage, and may also subsume other effects not matched by the plane-wave model and thus cannot be represented by a finite sum of plane waves (MPCs) \cite{jiang2022survey}. Its contribution to the \ac{DDIR} is thus a continuous function in the delay/angle domain. Based on the central limit theorem, it is modeled as a zero-mean complex Gaussian process with specific multi-dimensional covariance matrix; this implies the stipulation of a particular shape of the \ac{ADPS} of the cluster (a product of a one-sided single-exponential in delay with a Laplacian or \emph{von Mises} in angle), so that the estimation needs to extract only scalars, namely the decay constants in delay, \ac{DoA}, and \ac{DoD}.

As outlined in Sec. I.A, prior measurement campaigns have established that the DMC exhibits a clustered structure as well. In this work, we improve on this insight by considering the association between diffuse and specular clusters. In particular, we can split the diffuse clusters into two categories: (i) those associated with \ac{MPC} clusters, and (ii) standalone clusters. The former describe, e.g., the diffuse scattering from a surface with a roughness that is sufficiently small that both specular and diffuse reflections occur, while the latter could be created, e.g., by a (group of) bushes that have no trunks large enough to cause a specular reflection, but whose many leaves create significant diffuse scattering. Furthermore, diffuse clusters may appear in the extracted results when the discrete \acp{MPC} associated with a diffuse cluster are too weak to be detected (i.e., below the dynamic range). A full distinction between such algorithmic artifacts and truly physical standalone diffuse clusters is not possible in a noisy environment. More generally, the distinction between diffuse and specular components, in particular in the presence of noise, can never be completely unambiguous \cite[Sec. II.B]{jiang2022survey}.

\vspace{-10pt}
\subsection{Robust HRPE Based on RiMAX}

For the joint estimation of the \ac{MPC} and \ac{DMC} parameters, we use the RiMAX algorithm \cite{richter2005estimation}, which is the only one among the widely used \ac{HRPE} algorithms that explicitly includes the \ac{DMC}. At a high level, it iterates between estimation of the \acp{MPC} and of the DMC parameters, keeping the DMC and MPC parameters from the previous round, respectively. The algorithm is initialized with a CLEAN (serial interference cancellation) estimation, and optimization is performed with a gradient-based search of the parameters. \Acp{MPC} may be eliminated during the iteration process when their variance becomes too high; this together with the modeling of the \acp{MPC} prevents an overestimation of the number of discrete \acp{MPC} and controls model complexity. Since the minimum (taken over all location pairs) SNR of our measurements was 35 dB, all measurements were thresholded (at the cluster level) to provide this same dynamic range, enabling fair comparison \cite{cheng2025stochastic}. However, MPCs estimated by RiMAX are not subjected to additional thresholding, since introducing thresholding within the RiMAX framework would bias the estimation process and may prevent convergence to the optimal solution.

In \cite{cheng2024multi}, we extended the RiMAX framework to support multiple DMC clusters that have independent characteristics in the delay/angle domain. The DMC parameters are initialized from the total channel response, leading to a natural association between MPC clusters and their diffuse counterparts. To control the complexity of the extraction, the DMC cluster parameters in each iteration round are estimated serially in the delay domain, and the resulting DMC contributions are subtracted from the total signal; since the DMC parameters are refined in multiple iteration rounds, this serial approach does not detract from the validity of the final DMC parameters. The output of the algorithm is a set of specular (non-clustered) \acp{MPC} and an independent set of multi-cluster \ac{DMC} parameters describing the complete channel structure.

\vspace{-10pt}

\subsection{Clustering}

We first introduce the concept of the {\em \ac{MPC} distance} as a distance measure between pairs of components. 
To compute it, we first normalize the delay distance by the maximum cluster \ac{RMS} delay spread, $S_{\tau, \rm max}$, which in our case is $16.7$ns (corresponding to 5m), as obtained from Fourier analysis \cite{cheng2025stochastic} (clustering results for Fourier and HRPE are generally somewhat different; however, maximum cluster extent is of the same order): 
   \begin{align}
        \widetilde{\Delta \tau}(\tau_1,\tau_2) \triangleq |\tau_1 - \tau_2| / {S_{\tau, \rm max}}.
    \end{align}
    Similarly, the DoA difference is normalized to maximum angular spread 
    \begin{align}\label{eq:angle_dist}
        \widetilde{\Delta \phi}^2(\phi_1,\phi_2) \triangleq {|e^{j \phi_1} - e^{j \phi_2}|^2}/{S_{\phi, \rm max}^2},
    \end{align}
    where $S_{\phi, \rm max}^2$ denotes the maximum intra-cluster angular spread obtained from Fourier analysis. The unit phasor formulation in (\ref{eq:angle_dist}) is used to avoid issues with the periodicity of the angular spectrum, similar to the approach of Fleury for angular spread \cite{fleury2000first}. 
Finally, the aggregate MPC distance is computed as
    \begin{align}
        d_{\rm MPC} \triangleq \sqrt{\widetilde{\Delta \tau}^2 + \widetilde{\Delta \phi}_{\rm DoA}^2 + \widetilde{\Delta \phi}_{\rm DoD}^2},
    \end{align}
which provides a unitless, balanced metric for the assessment of clustering in the delay domain.

The clustering of MPCs and DMC then proceeds as follows:
for clustering \acp{MPC}, we adopt a reference-based approach. The estimated \ac{DMC} parameters\textemdash namely the starting delay, \ac{DoA}, and \ac{DoD} of each \ac{DMC} cluster\textemdash serve as reference centroids for distance evaluation. Each \ac{MPC} is then assigned to the \ac{DMC} cluster with the shortest \ac{MPC} distance. This differs from our earlier Fourier analysis paper \cite{cheng2025stochastic}, which employed multi-level \ac{KPD} in an unsupervised manner without predefined reference clusters. The present strategy is motivated by the inherent \ac{MPC}-\ac{DMC} association structure: since the associated \acp{DMC} constitute a subset of the full \ac{DMC} ensemble, they provide the natural references for \ac{MPC}-to-cluster assignment. Any remaining \ac{DMC} clusters without assigned \acp{MPC} are classified as standalone \ac{DMC}.

\vspace{-10pt}
\section{Stochastic Modeling of Channel Parameters}

Using the methodology in Sec. III, we extracted and clustered the MPCs and DMC from 21 \ac{LoS} and 17 \ac{NLoS} outdoor location pairs of the outdoor measurement campaign described in Sec. II. In this section, we describe the intra- and inter-cluster statistics over the ensemble of measurements. These statistics cover the number of clusters, number of \acp{MPC} in a cluster, the cluster parameter centroids\textemdash cluster arrival time, \ac{DoA} and \ac{DoD} centroids\textemdash , intra-cluster \ac{MPC} densities\textemdash excess delay and angular offsets\textemdash , intra-cluster power decay, intra-cluster statistics\textemdash delay spread, \ac{DoA} and \ac{DoD} angular spread\textemdash , cluster power, and their fitted distributions. Together, they constitute a complete modeling of D2D sub-THz channels. Note, however, that\textemdash like any measurement-based statistical model\textemdash our model is only applicable within the range of the underlying measurements (e.g., $d\le 100$m, no moving scatterers).

We analyze the cluster statistics separately for LoS and NLoS scenarios. Furthermore, in light of the fundamental difference in the underlying propagation mechanism, the cluster containing the LoS component (henceforth called the {\em direct} cluster) requires separate modeling. Thus, we distinguish three cases: (i) direct cluster (the cluster containing the direct/\ac{LoS} path), (ii) LoS clusters (these are clusters occurring in LoS scenarios but {\em not} containing the direct path), and (iii) NLoS clusters (i.e., clusters occurring in NLoS scenarios). We furthermore distinguish two {\em types} of clusters, namely clusters with both \ac{MPC} and \ac{DMC} (henceforth simply called ``MPC clusters''), and standalone DMCs (no associated \acp{MPC}).

To characterize the statistics of the extracted channel parameters, standard probability distributions are adopted, based on goodness-of-fit analysis. As an evaluation metric, we use the \ac{CvM} distance:
\begin{equation}
    {D}_{\mathrm{CvM}} 
    =\int_{-\infty}^{+\infty} 
    \left( \hat{F}(x) - F(x) \right)^2 \, \mathrm{d}F(x),
\end{equation}
where $F(x)$ denotes the empirical \ac{CDF} of the measured data, and $\hat{F} (x)$ denotes either the fitted \ac{CDF} or the empirical \ac{CDF} obtained from generated realizations of our postulated model.
The \ac{CvM} distance thus quantifies the integrated squared deviation between the two \acp{CDF} over the considered parameter range. The \acp{PDF} of the distributions used for the functional fits in our model are summarized in Table~\ref{tab:distributions}. For clarity, each distribution is denoted using compact notation henceforth. When continuous distributions are fitted for discrete-valued parameters (e.g., Gamma distribution for number of clusters), generation of parameter realizations implicitly uses rounding to the nearest integer.

\begin{table}[tb]
\centering
\caption{Standard Distributions Used in the Model}
\label{tab:distributions}
\begin{tabular}{ll}
\toprule
\textbf{Distribution} & \textbf{Probability Density Function} \\
\midrule
Gamma, $\mathcal{G}(a,b)$ 
& $f(x|a,b) = \frac{1}{b^a \Gamma(a)}x^{a-1}e^{-\frac{x}{b}}$ \\[4pt]

Normal, $\mathcal{N}(\mu,\sigma)$ 
& $f(x|\mu,\sigma)=\frac{1}{\sqrt{2\pi\sigma^2}}
e^{-\frac{(x-\mu)^2}{2\sigma^2}}$ \\[4pt]

Log-normal, $\text{LogN}(\mu,\sigma)$
& $f(x|\mu,\sigma) = \frac{1}{x\sigma \sqrt{2\pi}} e^{-\frac{(\ln{x} - \mu)^2}{2\sigma^2}}$ \\[4pt]

Laplace, $\mathcal{L}(\mu,\lambda)$
& $f(x|\mu,\lambda)=\frac{1}{2\lambda}
e^{-\frac{|x-\mu|}{\lambda}}$ \\ [4pt]

Bimodal Laplace, $\mathcal{BL}(w,\mu,\lambda)$ 
& $f(x|w,\mu,\lambda) = \sum_i \frac{w_i}{2\lambda} e^{-\frac{|x-\mu_i|}{\lambda_i}}$  \\ [4pt]

Rice, $\mathcal{R}(\nu,\sigma)$
& $f(x|\nu,\sigma) = \frac{x}{\sigma^2} e^{-\frac{x^2+\nu^2}{2\sigma^2}} I_0\left( \frac{x\nu}{\sigma} \right)$ \\[4pt]

Uniform, $U(a,b)$ 
& $f(x|a,b) = \frac{1}{b-a}$ \\
\bottomrule
\end{tabular}
\vspace{-10pt}
\end{table}

\vspace{-10pt}
\subsection{Number of Clusters}

\begin{figure}[tb]
\centering
\includegraphics[width=.7\linewidth]{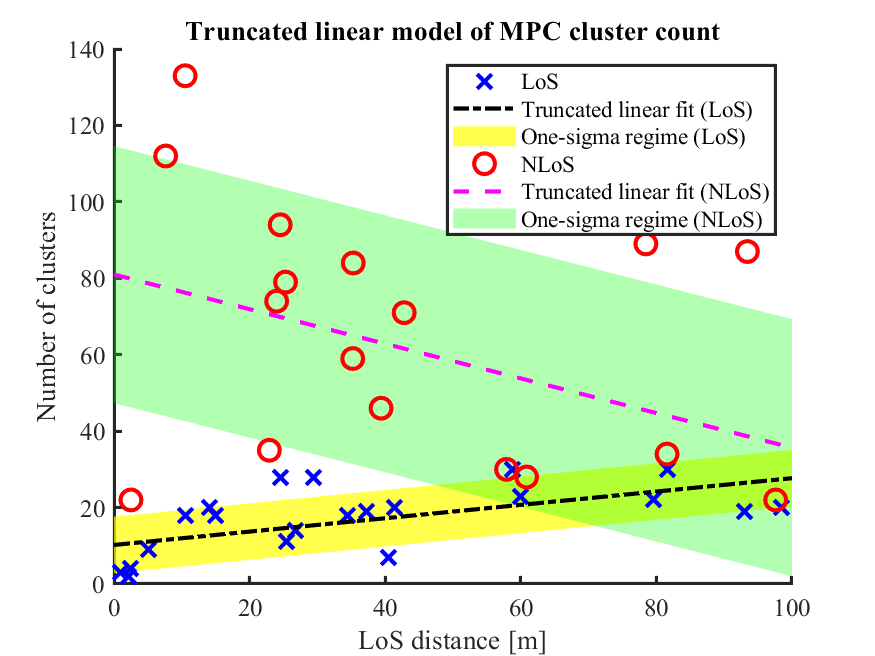}
\caption{Number of MPC clusters versus Tx–Rx distance for LoS and NLoS scenarios.}
\label{fig:Number_clusters_LoS}
\vspace{-16pt}
\end{figure}

We first analyze the number of \ac{MPC} clusters in \ac{LoS} and \ac{NLoS} scenarios. We observe a clear distance dependence in both scenarios, see Fig.~\ref{fig:Number_clusters_LoS}. The number of clusters in the \ac{LoS} case {\em increases} approximately linearly with the separation distance, (95\% confidence interval of the slope is [0.06, 0.29]). This can be explained by the fact that for short Tx-Rx distances, the ratio of runlength for reflected components versus the direct one is large for most clusters, leading to a relative strength
below the considered dynamic range. For \ac{NLoS} scenarios, the number of clusters seems to exhibit a decreasing trend with increasing distance, but the confidence interval of the slope encompasses both positive and negative values [-1.09, 0.19] so that any physical interpretation should be made with care. 

Motivated by Fig.~\ref{fig:Number_clusters_LoS}, we model the number of clusters as a linear function of distance. Specifically, the \ac{MPC} cluster count at Euclidean Tx-RX distance $d_i$, written as $N_i(d_i)$, is
\begin{align}\label{eq:cluster_number}
    N_i(d_i;\alpha,\beta,\sigma) = \alpha d_i + \beta + \epsilon(\sigma),
\end{align}
where $\alpha$ and $\beta$ denote the slope and intercept of the linear mean cluster count model, respectively, and the random variable $\epsilon(\sigma)$ captures the deviation from the mean, after which the resulting cluster count is rounded to the nearest integer. This formulation is applied separately to \ac{LoS} and \ac{NLoS} scenarios, resulting in different parameter sets that reflect the underlying propagation conditions.
The deviation is modeled as a {\em truncated} Gaussian, i.e., $\epsilon(\sigma)\sim \mathcal{N}(0,\sigma^2)$ subject to $N_i(d_i) \geq n_{\rm thresh}=1$, to account for the fact that there has to be at least one cluster, and realizations violating this constraint are discarded and re-generated. Consequently, we need to use a truncated \ac{MLE} for the estimation of the parameters $\alpha$, $\beta$, and $\sigma$, i.e., we maximize the truncated log-likelihood function
\begin{align}
    &\text{LLF}(N_i; \alpha, \beta, \sigma, n_{\rm thresh}) = & \\
    &-\sum_i \left[ \ln\sigma + \frac{(N_i - \mu_i)^2}{2\sigma^2} + \ln{(1 - \Phi\left( \frac{n_{\rm thresh}-\mu_i}{\sigma} \right) )}
    \right], \notag
\end{align}
where $\mu_i = \alpha d_i + \beta$ is the mean cluster count, and $\Phi(\cdot)$ denotes the standard normal \ac{CDF}. This formulation ensures consistency with the lower-bounded observations while properly accounting for the bias introduced by truncation. Note that for the numerical optimization, asymptotic expressions for the tail probability and the resulting gradients might have to be used to avoid numerical problems.


The estimated parameters for both the MPC and the standalone DMC (sDMC) clusters are summarized in Table~\ref{tab:summary}. 
The goodness-of-fit is evaluated by the CvM distance between the measured cluster counts (using the ensemble of observations at the different measured distances) and the synthetically generated realizations (assuming the same distribution of distances), yielding values of 0.041 (MPC LoS), 0.008 (sDMC LoS), 0.004 (MPC NLoS), and 0.034 (sDMC NLoS), confirming a close agreement between the model and measurements.

\vspace{-10pt}

\subsection{Number of MPCs per Cluster}

\begin{figure}[tb]
\centering
\includegraphics[width=.7\linewidth]{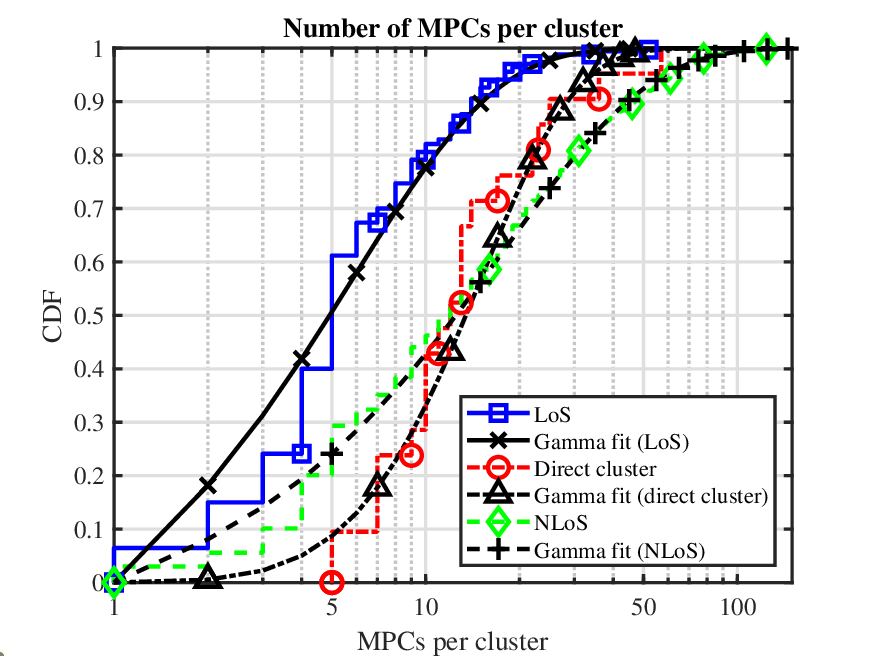}
\caption{The number of \acp{MPC} in a cluster.}
\label{fig:Number_MPC_per_cluster}
\vspace{-14pt}
\end{figure}

Fig.~\ref{fig:Number_MPC_per_cluster} shows the empirical \ac{CDF} and the fit for the number of MPCs per cluster. Notably, while the Poisson distribution is widely used for the number of paths, it does not show good agreement with our results, which are much more heavy-tailed, indicating a deviation of the environmental geometry (scatterers of a cluster distributed according to a Poisson point process) that underlies the existing models. In our measurements, a Gamma distribution provides a good fit, though with different parameters for the direct cluster, LoS clusters, and NLoS clusters. The LoS clusters have a noticeably smaller number of MPCs (mean value of 6) compared to the other two cases (14.5 and 21.9 for direct and NLoS clusters, respectively), which can be partially attributed to their smaller dynamic range. We also note a larger shape parameter (i.e., a wider spread of the \ac{CDF}) in the NLoS case, reflecting the more complex propagation environment, including the possibility of an increased number of MPCs due to interactions with more objects compared to clusters in LoS scenarios. The shape and scale parameters of all the fits are provided in Table~\ref{tab:summary}. 

\vspace{-10pt}
\subsection{MPC Cluster Arrival Time}

\begin{figure}[tb]
\centering
\includegraphics[width=.7\linewidth]{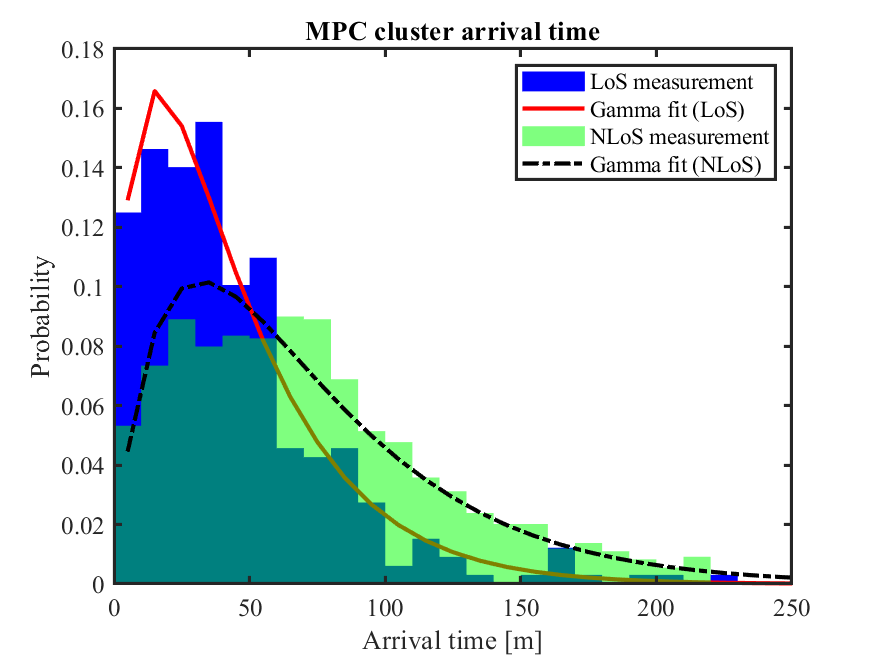}
\caption{The MPC cluster arrival time relative to \ac{LoS} distance.}
\label{fig:cluster_arrival_time}
\vspace{-18pt}
\end{figure}

We define the cluster arrival times  as the delay relative to the runtime of the direct component (no matter whether it physically exists, as in the LoS case, or not, as in NLoS environments) and the smallest-delay MPC of a cluster. This normalization guarantees non-negativity of the arrival time, and furthermore removes distance-dependent effects, so that all location pairs can be included fairly in the statistics. Fig.~\ref{fig:cluster_arrival_time} shows the histogram (empirical \ac{PDF}) of the arrival time (expressed in meter, i.e., multiplied with $c_0=3\cdot 10^8)$), and the best fit with a Gamma distribution. 
Although the Poisson assumption is widely adopted in classical clustered channel models, such as S-V framework, for our measurements it fails to represent the observed pronounced heavy-tail characteristics, significantly underestimating the probability mass at larger arrival times. In contrast, the Gamma distribution provides a substantially improved fit across the full range of observed values.
Not shown is the arrival time of the direct cluster, which is zero by definition. The LoS clusters show a higher concentration of the arrival time near zero than the NLoS clusters (Gamma fits with $a = 1.53$ and $b = 28.71$ for LoS and $a = 1.81$ and $b = 39.31$ for NLoS, respectively). Again, this is partly attributable to the different dynamic ranges in LoS (when the direct cluster is excluded) and NLoS cases, and partly to the more complex environment in NLoS situations, where we conjecture that longer detours may still provide appreciable received power, e.g., due to waveguiding.

We finally note that the estimated Gamma parameters differ from those reported in the Fourier-based modeling framework \cite{cheng2025stochastic}. Although the same statistical distribution, i.e., Gamma, is employed, the numerical parameters are not identical across modeling works. These discrepancies reflect differences in the modeling approach, in particular the separation of MPC and DMC clusters in the current paper (which is not possible for Fourier evaluations), rather than indicating inconsistencies. 

\vspace{-10pt}
\subsection{MPC Cluster DoA/DoD Centroids}

\begin{figure}[tb]
\centering
\includegraphics[width=.7\linewidth]{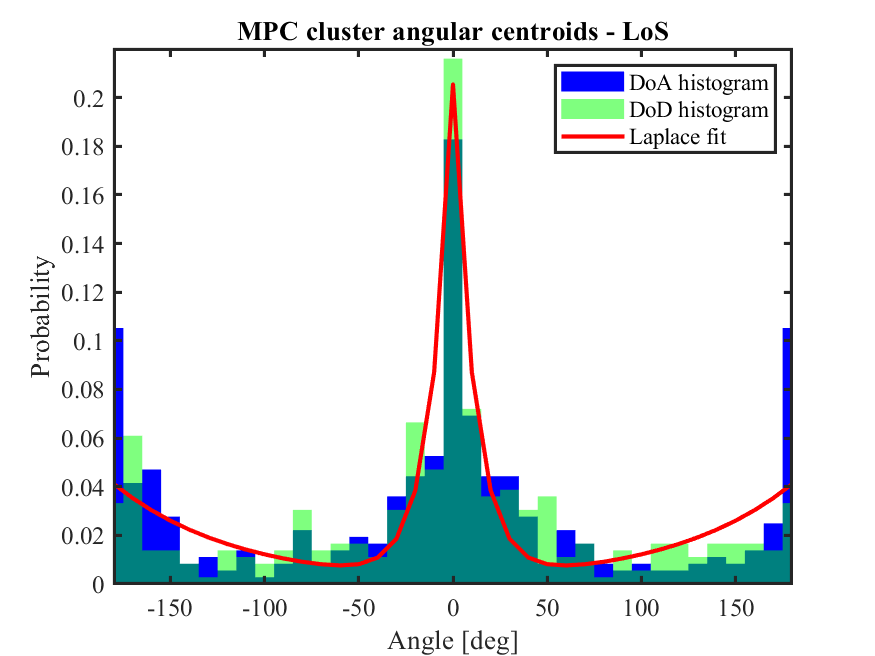}
\caption{The \ac{MPC} cluster angular centroids relative to \ac{LoS} angular direction under \ac{LoS} condition.}
\label{fig:cluster_centroid_LoS}
\vspace{-12pt}
\end{figure}

The cluster \ac{DoA}/\ac{DoD} centroids indicate the power-weighted mean angles of a cluster. The origin of the angular coordinate system ($0 ^{\circ}$) is defined as the boresight direction between \ac{Tx} and \ac{Rx} for both \ac{LoS} and \ac{NLoS} scenarios, though in \ac{NLoS} the boresight connection (direct path) is blocked. 
Since the environments of Tx and Rx are statistically equivalent (same height of nodes above ground), we can expect similar statistical characteristics of \ac{DoA} and \ac{DoD} across LoS and NLoS conditions, which is confirmed by Fig.~\ref{fig:cluster_centroid_LoS}. 
Thus, we jointly evaluate \ac{DoA} and \ac{DoD} and interpret them as realizations of the same angular distribution. This ensures consistency as well as improved statistical robustness. Analogously, \ac{DoA} and \ac{DoD} are jointly treated in all subsequent angle-related analysis, including inter-cluster statistics of standalone \ac{DMC} and intra-cluster decay parameters for \acp{MPC} and \acp{DMC}. Note that this approach does {\em not} hold for links where one link end is at a different height than the other.

The angular distribution of clusters is often modeled as a Laplacian or Gaussian distribution. However, results from our measurements show significant differences. As visible in Fig.~\ref{fig:cluster_centroid_LoS}, in the LoS case, the cluster centers show two peaks centered at $0^{\circ}$ and $180^{\circ}$. This is due to pronounced reflections behind Tx and Rx, respectively. Consequently, the \ac{PDF} of the cluster centers is better described by a {\em bimodal} Laplacian distribution, as also shown in Fig.~\ref{fig:cluster_centroid_LoS}.

The situation is different in NLoS situations. There, dominant directions are determined by the presence of suitably located reflectors (in the quad environment) or waveguiding (in the street canyon environment). These directions are essentially independent of the direction of the (blocked) direct path. Consequently, we can anticipate an (almost) uniform distribution of the cluster center angles. This conjecture is confirmed 
by the small \ac{CvM} distance of approximately 0.003 for a uniform \ac{PDF} with respect to the measured distribution.


\vspace{-12pt}
\subsection{Intra-Cluster MPC Excess delay}

One of the key aspects of our model is the stochastic description of intra-cluster MPC arrival time and angle. The 3GPP model assumes equal arrival time and fixed offsets from the mean angle. Fourier-based models describe only the average ADPS and implicitly assume equal spacing in delay and angle domains for the MPCs. S-V-type models assume Poisson arrival times, corresponding to exponential inter-arrival times of the MPCs. Since our measurements show different behavior, we need to provide a new model.

\begin{figure}[tb]
\centering
\includegraphics[width=0.8\linewidth]{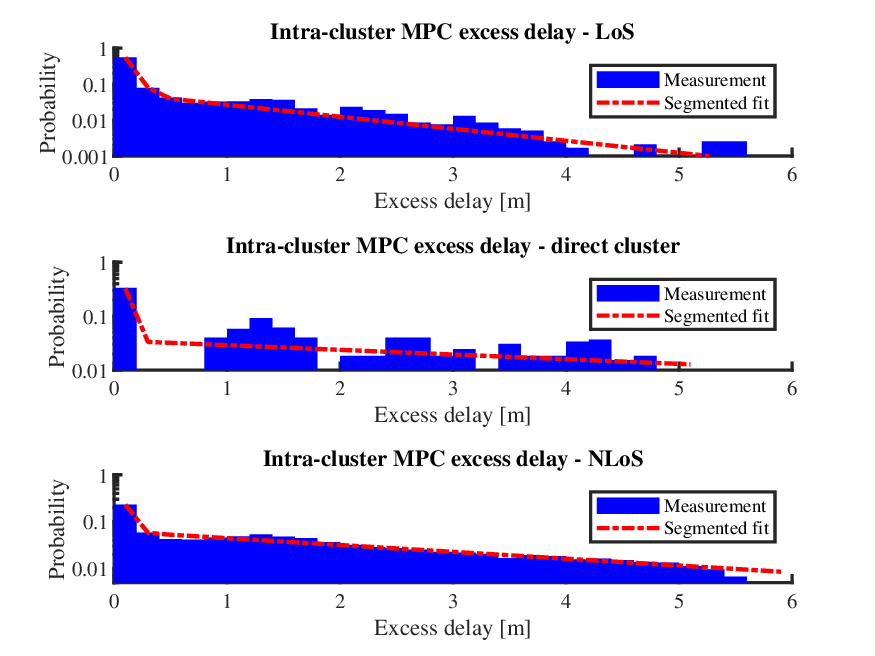}
\caption{Measured \ac{MPC} intra-cluster excess delay under \ac{LoS}, \ac{LoS}-cluster, and \ac{NLoS} conditions.}
\label{fig:MPC_intra_excessdelay}
\vspace{-14pt}
\end{figure}

We observe, for all cluster types a sharp peak of the arrival time near delay zero (all delays are normalized to the arrival of the first MPC in the cluster), followed by a slower decay; see Fig.~\ref{fig:MPC_intra_excessdelay}. Thus, the model is chosen to provide a \ac{BP} $\tau_{\rm BP}=0.3$m,\footnote{For easier interpretability, we given the excess ``delay'' (runlength) in meters; the excess delay in seconds is obtained by division by $c_0=3 \times 10^8$.} i.e., to separately model the arrival time for small ($\le \tau_{\rm BP}$ excess delay) and large (in the range $(\tau_{\rm BP}, 6]{\rm m}$, where 6m is the maximum delay observed in our analysis): 
    \begin{align}\label{eq:segmented_fit}
    f(x|\tau_{\rm BP},\lambda_1,\lambda_2) = \left\{ \begin{matrix}
        C e^{-\lambda_1 x} &x \leq \tau_{\rm BP} \\
        C_1 e^{-\lambda_2 (x-\tau_{\rm BP})} & \tau_{\rm BP} < x \leq 6
    \end{matrix} \right.
    \end{align}
    where $C_1=C e^{-\lambda_1 \tau_{\rm BP}}$ and the constant $C$ is chosen to satisfy the normalization condition $C \int_{0}^{\tau_{\rm BP}} e^{-\lambda_1 x} dx + C_1 \int_{\tau_{\rm BP}}^{6} e^{-\lambda_2 (x-\tau_{\rm BP})}dx = 1$.
This segmentation provides a better tradeoff between modeling fidelity and complexity of the model. The fitting parameters $\lambda_1, \lambda_2$ are given in Table II. 

\vspace{-12pt}

\subsection{Intra-Cluster MPC DoA/DoD Offset}

\begin{figure}[tb]
\centering
\includegraphics[width=\linewidth]{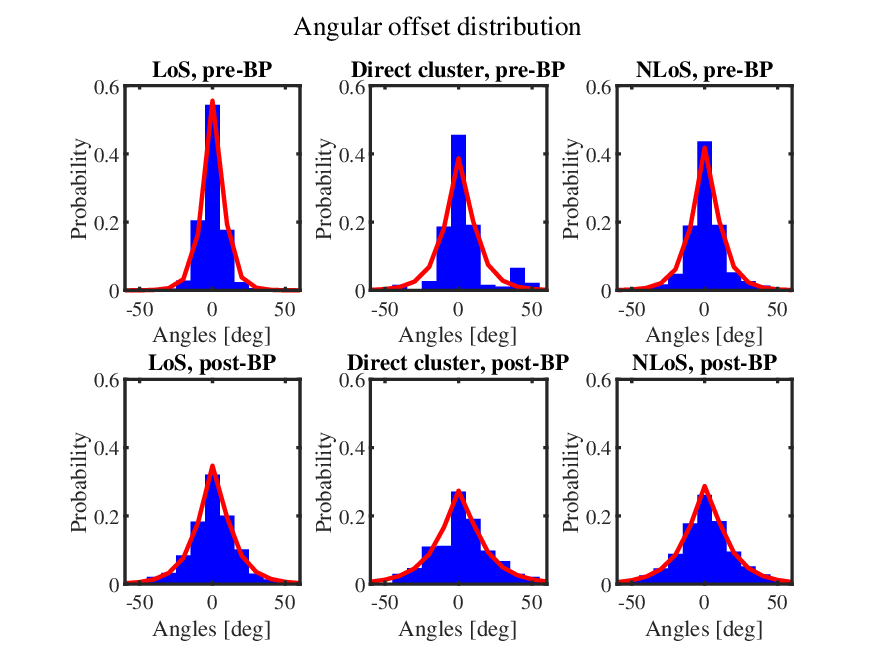}
\caption{Distribution of MPC angular offsets from cluster centroids under LoS, direct-cluster, and NLoS conditions, in pre- (top) and post- (bottom) \ac{BP} regions. Blue bars: measured histograms. Red solid lines: fitted Laplacian.}
\label{fig:DoA_DoD_offset}
\vspace{-12pt}
\end{figure}

The angles of the MPCs are modeled as relative offsets from the cluster centroid angles. These offsets exhibit a symmetric, zero-centered distribution across all propagation conditions, see Fig.~\ref{fig:DoA_DoD_offset}. This is in line with expectations, as there is no physical effect introducing angular asymmetries. 

Again, the MPCs show significantly different behavior in the two regions $\tau \le \tau_{\rm BP}$ and $\tau > \tau_{\rm BP}$. In the former case, the offset angles are more tightly concentrated near 0 than the latter; 
this introduces a coupling between delay and angular offset, whereas within each of the two regions, angle-delay correlation is negligible (Pearson correlation coefficient of 0.03). In each case, the distribution function of the offset is a Laplacian; however, no measured MPCs exceeded $60^{\circ}$ offset across all campaigns. Again, the parameters of the distribution, $\lambda^{\rm pre}$ and $\lambda^{\rm post}$ are given in Table~\ref{tab:summary} for the three different cluster types. We find that the LoS clusters have somewhat smaller width than the direct and the NLoS clusters, again related to the dynamic range issue mentioned previously.

\vspace{-10pt}
\subsection{LoS Component Rician Fading}

In a LoS environment, a direct component exists, and its parameters (delay, angle) can be computed {\em deterministically} from the location of Tx and Rx. However, while the power of the direct component is theoretically also deterministically given (from Friis' equation), the values extracted via RiMAX show residual deviations from that theoretical value. This is due to the fact that even HRPE does not have infinite resolution at finite SNR; thus, low-amplitude MPCs that originate close to the Tx or Rx are merged with the direct component. 

We consequently model the power variations relative to the nominal (free-space) value, postulating that the amplitude has a Rician \ac{PDF}. We create an ensemble of amplitudes (normalized by their free-space values) from the different LoS points, and use a \ac{ML} estimator for the LoS parameters. In particular, we construct the (negative) log-likelihood function of the Rician distribution and minimize it using constrained nonlinear programming (interior-point algorithm), with positivity constraints imposed on both parameters to ensure physical validity. Initial parameter values are chosen based on empirical envelope statistics, i.e., $\nu$ is initialized as the maximum magnitude, and $\sigma$ is initialized by \ac{std} of all magnitude data. This results in magnitude and variance estimate of  $\hat{\nu} = 0.91$ and $\hat{\sigma} = 0.25$, respectively. The estimated $\hat{\nu}<1$ indicates that, on average, the extracted \ac{LoS} component is weaker than the nominal free-space prediction by about 0.5 dB. This deviation can be attributed to inevitable calibration imperfections and the finite-resolution limitations of the RiMAX-based \ac{HRPE}, which prevent a ``clean'' extraction. Consequently, part of the deterministic \ac{LoS} energy may be partially redistributed during parameter estimation. 

\vspace{-12pt}
\subsection{MPC Intra-Cluster Power Decay}

The fraction of cluster power carried by the individual MPCs exhibits a distinct dependence on the delay and angle offsets from the cluster centroid, $\Delta \tau$ and $\Delta \phi$. The measured data reveal a breakpoint in the delay domain, separating two decay regimes. Accordingly, exponential decay models are fitted independently before and after the breakpoint. In each regime, the power decay is well approximated by
\begin{equation}\label{eq:decay_model}
    p(\Delta \tau,\Delta \phi_{\rm R}, \Delta \phi_{\rm T}) = e^{\zeta} e^{-\lambda_{\tau} \Delta \tau} e^{-\lambda_{\phi} |\Delta \phi_{\rm R}|} e^{-\lambda_{\phi} |\Delta \phi_{\rm T}|},
\end{equation}
where $\zeta$ represents the logarithmic power intercept, and $\lambda_{\tau}$ and $\lambda_{\phi}$ denote the decay constants governing the power attenuation with delay and angular offsets, respectively. 
The fitting is influenced by more data being available for small offsets. To compensate for this effect, the fitting is weighted, with the weights inversely proportional to the data density, similar to the basic principle in \cite{karttunen2016path}. Unlike the bin-based approach with explicit upper limits adopted therein, our formulation employs a continuous density function defined over bounded excess delay and angular offset, such that the density (and thus its inverse weighting) remains naturally limited. 

Due to the limited dynamic range, a least-square fit (on a dB scale) of the decay constant would result in a biased estimate (again similar to the observations in \cite{karttunen2016path}). To resolve this problem, we adopt an ML estimator for the truncated (censored) data that explicitly accounts for the detection threshold, and thus produces statistically consistent decay parameter estimates, similar in spirit to (7).

\vspace{-12pt}
\subsection{MPC Phase}

Besides the amplitude, we also need to determine the phase of the MPCs. From physical considerations, we conjecture that it should follow a uniform distribution\textemdash both the propagation processes (e.g., phase shift during a reflection), and the differences in runlength tend to randomize the phase. The analysis of the histogram of the measurements confirms this conjecture. The goodness-of-fit analysis yields a \ac{CvM} distance of 0.015 with respect to the uniform model, indicating good agreement between the empirical data and the uniform distribution. 

\vspace{-12pt}

\subsection{Standalone DMC Inter-cluster Statistics }

We now turn to the standalone DMC clusters. In terms of the inter-cluster characteristics, we need to describe similar parameters as for the MPC clusters, namely the cluster arrival time, and the distribution of the DoA/DoD centroids. For both characteristics, the distributions have similar shapes as those of the MPC clusters (for this reason, we do not show their plots), though they have different parameters. In particular the arrival time follows a Gamma distribution with parameters $a = 2.78$ and $b = 31.33$ for \ac{LoS} scenarios, and $a = 4.58$ and $b = 27.29$ for \ac{NLoS} scenarios. This indicates that NLoS scenarios lead to a higher probability of standalone clusters at longer distances, which is in line with intuition.

In the LoS scenarios, the distribution of the angular centroids shows two distinct angular modes that can be well approximated by a bimodal Laplacian distribution, in agreement with the distribution for the MPC clusters. The location (mode) parameters remain centered at $0^{\circ}$ and $180^{\circ}$, consistent with the MPC cluster statistics. The corresponding scale parameters are of comparable magnitude, with values of 0.20 and 0.29 for the dominant modes, and 1.14 and 0.56 for the weaker mode, as provided in Table~\ref{tab:summary}.
For the NLoS scenarios, the angular distribution is well characterized by a uniform distribution between 0 and $2 \pi$.

\vspace{-12pt}
\subsection{Intra-cluster DMC Statistics}

For the intra-cluster statistics of the DMC, we model the double-directional delay power spectrum (DDDPS) of each cluster as the product of a single-exponential decay in delay, and Laplacian spectra in DoA and DoD (again assuming that the two link ends are statistically equivalent in terms of angles)
\begin{equation}
    DDDPS(\tau, \phi^{\rm R},\phi^{\rm T})=e^{-\tau/a}e^{-|\phi^{\rm R}|/b_{\rm R}} e^{-|\phi^{\rm T}|/b_{\rm T}}.
\end{equation}
Here, $a$ and $b_{\rm R/T}$ denote the decay constants in the delay and angular domains, respectively, and fully determine the intrinsic cluster profile, where the angular decay constants for \ac{DoA} and \ac{DoD} are modeled as independent and identically distributed realizations. In an ideal system with unlimited dynamic range, the decay constants admit simple closed-form relationships with the \ac{RMS} delay spread and angular spread. However, the measurements are subject to power truncation due to a finite dynamic range (35 dB in this work). This leads to a more complicated relationship between the cluster delay spread and the decay constants. Specifically, the mapping between the observed spreads and the decay constants depends on the effective dynamic range of each cluster, i.e., the ratio between the peak power and the truncation threshold. Using the analytical relationships derived in \cite{cheng2025stochastic}, the decay constants $a$ and $b$ are recovered such that the truncated DDDPS reproduces the measured delay and angular spreads.

\begin{figure}[H]
\centering
\includegraphics[width=0.8\linewidth]{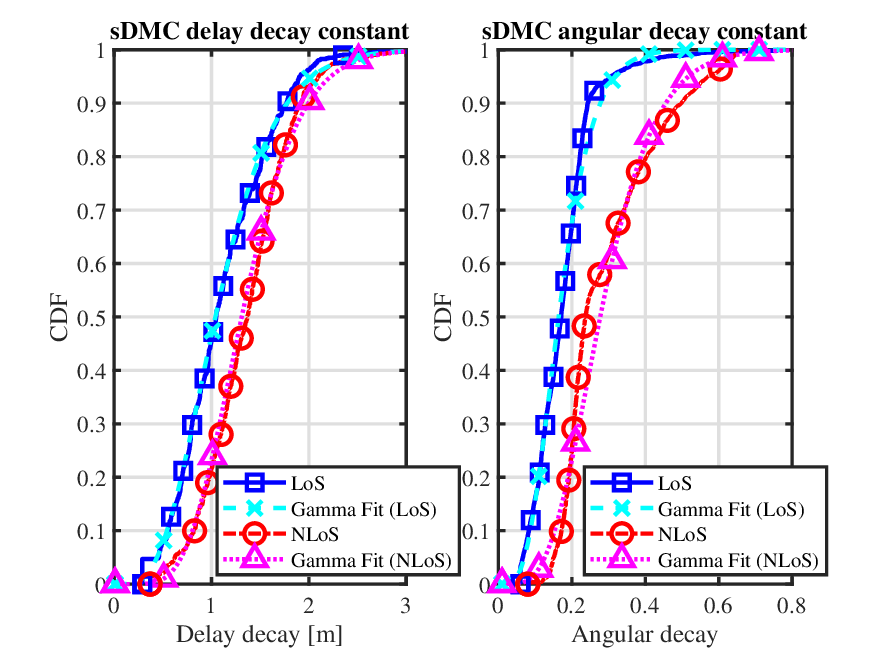}
\caption{Empirical \acp{CDF} of estimated standalone DMC (sDMC) delay (left) and angular (right) decay constants for \ac{LoS} and \ac{NLoS} scenarios, and corresponding Gamma fits.}
\label{fig:standalone_DMC_DSAS_comp}
\vspace{-10pt}
\end{figure}

The extracted decay constants are then modeled statistically. While lognormal distributions are commonly used for cluster delay spreads in the literature, a \ac{CvM} distance showed better fits using Gamma distributions. The fitted distributions shown in Fig.~\ref{fig:standalone_DMC_DSAS_comp} correspond to the standalone DMC case, serving as a representative example. For associated DMC, a similar model is used; however, the resulting Gamma parameters differ. For brevity, the corresponding distributions are not illustrated, and only the fitted values are reported in Tables III and IV.

\vspace{-10pt}
\subsection{Cluster Power }

We now turn to the distribution of the cluster power. For this, we need to distinguish between MPC clusters (containing specular MPCs and DMC), and the standalone DMC clusters. The functional form of the distribution and its dependence on distance are assumed to be identical in both cases, but the parameterization is different. We first extract the total cluster power from the clustered RiMAX results as $P_c = P_c^{\rm MPC} + P_c^{\rm DMC}$ for the MPC clusters, and $P_c = P_c^{\rm DMC}$ for standalone DMC clusters.
We note that we here model 
 the total power of the MPC+DMC clusters together with the ratio of the contributions from the two types of paths, rather than 
 modeling MPC and DMC cluster power independently, because an direct (and thus more accurate) modeling of the total power is more important. Furthermore, it is more in line with the philosophy of Rimax, which ensures that the sum of MPC and DMC explains the total measured signal. 

In order to ensure a fair comparison, the cluster powers for a particular Tx-Rx location pair are normalized by the free-space pathloss valid for that distance (irrespective of whether the direct component actually exists). This normalized cluster power shows, on average, a decrease as a function of the arrival time. In particular, we found that it is well modeled as a function of the {\em ratio} of the cluster runlength and the direct-component runlength rather than the difference, as is assumed in some other papers. Furthermore, we observe random deviations of the cluster power from this deterministic behavior, which\textemdash on a dB scale\textemdash can be described by a zero-mean Gaussian distribution, whose variance also exhibits a distance dependence (while $\sigma$ can decrease with distance, it can of course never become negative). To summarize, the cluster power as a function of distance can be written as 
\begin{align}
    \tilde{P}_{c| {\rm dB}} &= \bar{P}(d,d_{\rm LoS}|\alpha_{\rm cl}, \beta_{\rm cl}) + \chi_{\sigma} \notag \\
    &= \alpha_{\rm cl} + 10\beta_{\rm cl} \log_{10}\!\frac{d}{d_{\rm LoS}} + \chi_{\sigma}, \chi_{\sigma} \sim \mathcal{N}(0,\sigma^2)\label{eq:mean_power}
\end{align}
\vspace{-1.5em}
\begin{align}
    \sigma(d,d_{\rm LoS}|\rho_{\rm cl}, \eta_{\rm cl}) = \rho_{\rm cl} + 10 \eta_{\rm cl} \log_{10}{\frac{d}{d_{\rm LoS}}}. \label{eq:power_std}
\end{align}

The fitting of this distribution to the measurement data is nontrivial because (i) the distance dependence of the shadowing variance prevents a simple least-square fitting, and (ii) the limited dynamic range and resulting thresholding of the clusters must be taken into account. Furthermore, the thresholding not only impacts the visible power of each cluster, but also the {\em number} of clusters that are visible at different distances. We solve this problem by constructing a suitable log-likelihood function for the truncated power, similar to \cite{cheng2025stochastic}. Thus, the likelihood of each observation is computed using the \ac{PDF} of the \emph{truncated normal distribution}
\begin{align}\label{eq:LLR-censored}
    &\text{LLF}(\alpha_{\rm cl},\beta_{\rm cl},\rho_{\rm cl},\eta_{\rm cl}) = \sum_i w_i\left[ -\ln\sigma(d_i,d_{\rm LoS})\right. - \\
    &\left.  \frac{(\text{PL}(d_i,d_{\rm LoS}) - \bar{P}(d_i,d_{\rm LoS}))^2}{2\sigma^2(d_i,d_{\rm LoS})} -\ln\Phi(\frac{\widetilde{\text{PL}} - \bar{P}(d_i,d_{\rm LoS})}{\sigma(d_i,d_{\rm LoS})}) \right]. \notag
\end{align}
The goal is to maximize the log-likelihood with respect to the model parameters, $(\alpha_{\rm cl}, \beta_{\rm cl}, \rho_{\rm cl}, \eta_{\rm cl})$, allowing for accurate estimation despite censoring.

To solve the resulting nonlinear minimization problem, we adopt an alternating optimization strategy. Since the four parameters are mutually coupled, joint estimation in closed form is not available. The iterations begin with the estimation of the mean power parameters, $\alpha_{\rm cl}$ and $\beta_{\rm cl}$, which are then fixed while estimating the variance-related parameters. These two steps are performed iteratively until convergence criteria\textemdash namely, that the relative updates to parameters fall below $10^{-5}$, or a maximum number of 1000 iterations\textemdash are fulfilled.

The parameter results both for the MPC+DMC clusters, and the DMC-only clusters are give in Table~\ref{tab:summary}.

\begin{figure}[tb]
    \centering
    \begin{subfigure}[t]{0.46\columnwidth}
        \centering
        \includegraphics[width=\linewidth]{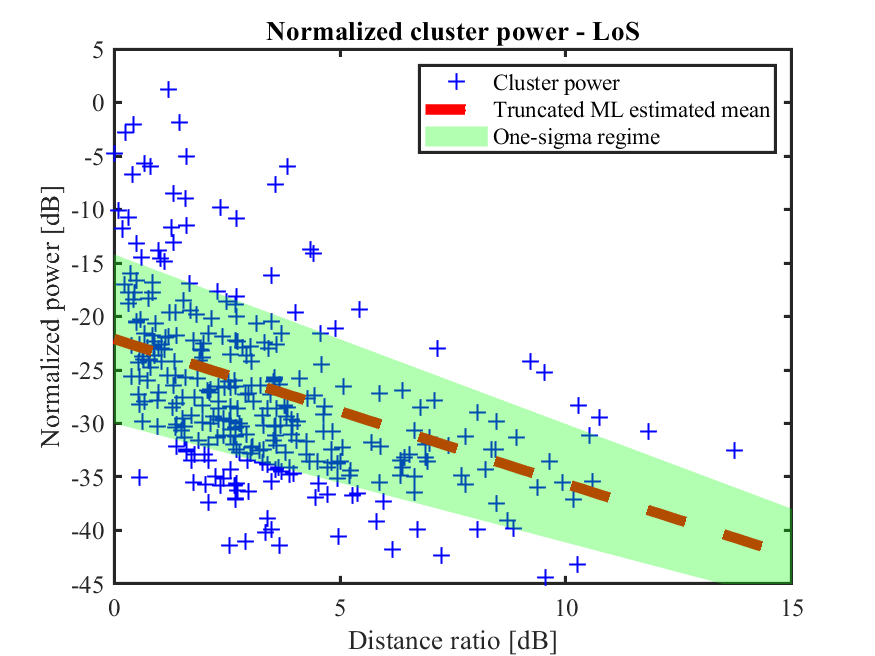}
        \caption{The total cluster power against the distance ratio in \ac{LoS} scenarios.}
        \label{fig:cluster_pwr_LoS}
    \end{subfigure}
    \begin{subfigure}[t]{0.46\columnwidth}
        \centering
        \includegraphics[width=\linewidth]{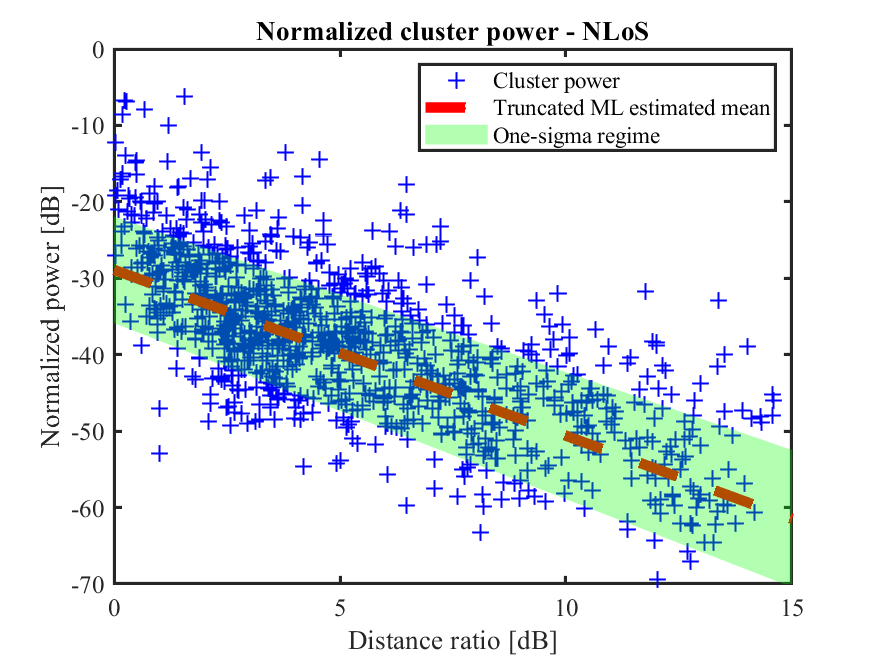}
        \caption{The total cluster power against the distance ratio in \ac{NLoS} scenarios.}
        \label{fig:cluster_pwr_NLoS}
    \end{subfigure}
    \\
    \centering
    \begin{subfigure}[t]{0.46\columnwidth}
        \centering
        \includegraphics[width=\linewidth]{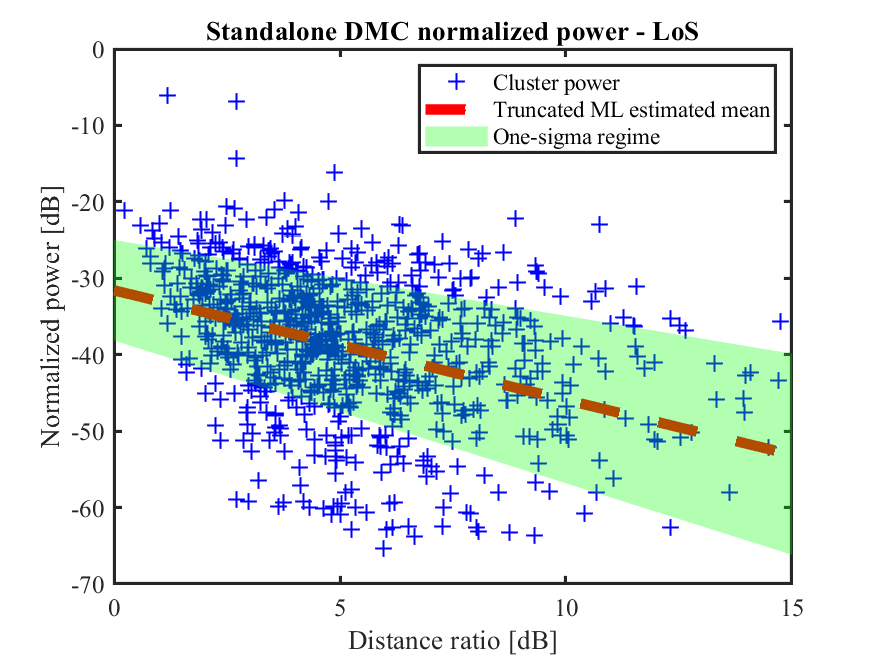}
        \caption{The standalone DMC power against the distance ratio in \ac{LoS} scenarios.}
        \label{fig:standalone_DMC_pwr_LoS}
    \end{subfigure}
    \begin{subfigure}[t]{0.46\columnwidth}
        \centering
        \includegraphics[width=\linewidth]{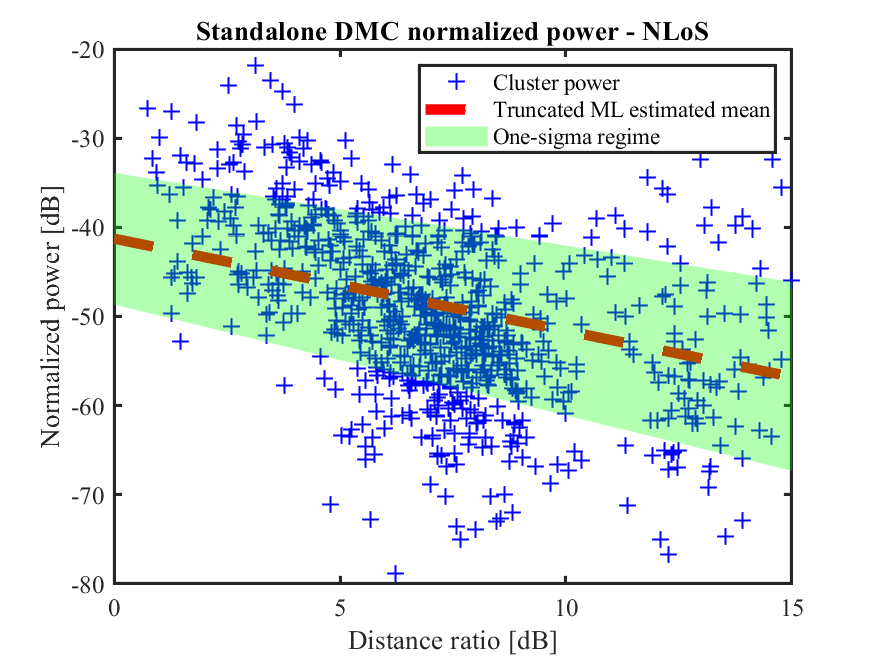}
        \caption{The standalone DMC power against the distance ratio in \ac{LoS} scenarios.}
        \label{fig:standalone_DMC_pwr_NLoS}
    \end{subfigure}
    \caption{Measured normalized power versus distance ratio for cluster under LoS and NLoS conditions: total cluster power (a)-(b) and standalone DMC power (c)-(d).}
    \label{fig:MPC_excess_delay}
    \vspace{-1em}
\end{figure}

\begin{figure}[H]
\centering
\includegraphics[width=.70\linewidth]{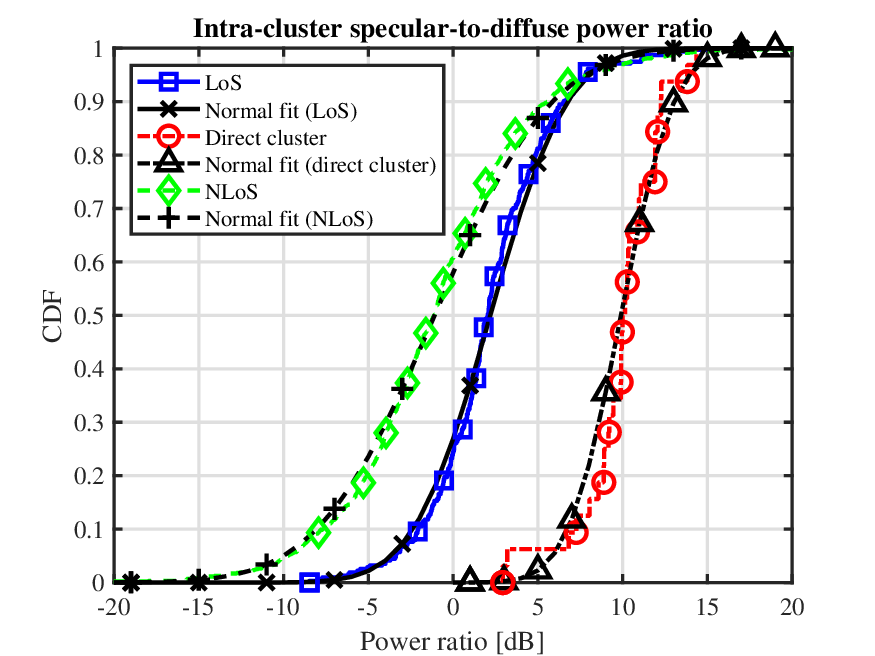}
\caption{The intra-cluster specular-to-diffuse power ratio under \ac{LoS}, \ac{LoS}-cluster, and \ac{NLoS} conditions.}
\label{fig:MPC_DMC_pwr_ratio}
\vspace{-10pt}
\end{figure}

\vspace{-16pt}
\subsection{Intra-Cluster Specular-to-Diffuse Power Ratio }

We now model the relative contribution of the DMC in the MPC+DMC clusters, namely the \ac{SDR}. Specifically, we consider the ratio of MPC to DMC power, which means the power in the specular components to those in the diffuse. This definition bears some formal similarity to the definition of the Rice factor (though the Rice factor assumes a {\em single} dominant component); this motivates to approximate its distribution similar to the distribution of the Rice factor common in the literature, namely 
as a normal distribution on a dB scale, i.e., $\text{SDR} = 10\log_{10}{{P^{\rm MPC}}/{P^{\rm DMC}}}$,
where $P^{\rm MPC}$ and $P^{\rm DMC}$ stand for total \ac{MPC} and \ac{DMC} power within cluster; the cluster index is omitted for brevity. More importantly, comparison of this distribution to the measurement results shows excellent agreement, as shown in Fig.~\ref{fig:MPC_DMC_pwr_ratio}. Furthermore, we see that for both LoS and NLoS clusters, a significant percentage of the energy is in the DMC (mean value of the SDR is 2.2 dB in LoS, and -1.1 dB in NLoS). Only the direct cluster shows a dominance of specular contributions (SDR=9.9 dB). This demonstrates again the importance of incorporating diffuse MPCs in the channel model.

\vspace{-10pt}
\subsection{Summary of statistical parameters }

Table~\ref{tab:summary}, \ref{tab:sDMC_spread}, and \ref{tab:aDMC_spread} provide a comprehensive summary of the statistical models and parameters extracted for sub-THz outdoor D2D channels. They provide the fitting equations (shape of the distribution, or dependence on distance), as well as the parameters of the fit. The parameters are given for (i) direct MPC clusters, (ii) LoS MPC clusters, and (iii) NLoS MPC clusters, as well as (iv) associated DMC clusters (whose parameters are considered to be the same for direct and LoS clusters), and (v) standalone DMC clusters. 



\begin{table*}[tb]
    \centering
    \caption{The summary of \ac{sub-THz} outdoor \ac{D2D} channel modeling}
    \begin{tabular}{|c|c|c|c|c|}
        \hline        
        &\textbf{\rule[-5pt]{0pt}{15pt} Modeling } &\textbf{Direct cluster} &\textbf{\ac{LoS}} &\textbf{\ac{NLoS}}\\
        \hline
        \parbox[c]{0.15\linewidth}{\vspace{8pt}\centering \textit{MPC cluster number}\vspace{8pt}} 
        &\parbox[c]{0.08\linewidth}{\centering (\ref{eq:cluster_number}) \\
        $(\alpha, \beta, \sigma)$
        }
        &1
        &\parbox[c]{0.15\linewidth}{\centering $\alpha=0.17$, $\beta=10.22$ \\
        $\sigma = 7.40$}
        &\parbox[c]{0.17\linewidth}{\centering $\alpha=-0.45$, $\beta=80.98$ \\
        $\sigma = 33.63$} 
        \\
        \hline
        \parbox[c]{0.15\linewidth}{\vspace{4pt}\centering \textit{Standalone DMC number} \vspace{4pt}} 
        &\parbox[c]{0.08\linewidth}{\centering (\ref{eq:cluster_number}) \\
        $(\alpha, \beta, \sigma)$ }
        &\textbf{--}
        &\parbox[c]{0.15\linewidth}{\centering $\alpha=0.21$, $\beta=39.56$ \\
        $\sigma = 17.10$}
        &\parbox[c]{0.17\linewidth}{\centering $\alpha=-0.82$, $\beta=77.83$ \\
        $\sigma = 31.77$}  \\
        \hline
        \parbox[c]{0.12\linewidth}{\vspace{4pt}\centering \textit{MPC number} \vspace{4pt}} 
        &\parbox[c]{0.08\linewidth}{\centering $\mathcal{G}(a,b)$ }
        &\parbox[c]{0.15\linewidth}{\centering $a=3.32$, $b=3.57$}
        &\parbox[c]{0.15\linewidth}{\centering $a=0.96$, $b=6.69$}
        &\parbox[c]{0.15\linewidth}{\centering $a=0.78$, $b=24.39$}  \\
        \hline
        \parbox[c]{0.12\linewidth}{\vspace{4pt}\centering \textit{\ac{MPC} cluster arrival time} $[\textbf{\text{m}}]$\vspace{4pt}} 
        &\parbox[c]{0.08\linewidth}{\centering $\mathcal{G}(a,b)$ }
        &\textbf{--}
        &\parbox[c]{0.15\linewidth}{\centering $a=1.47$, $b=29.85$}
        &\parbox[c]{0.15\linewidth}{\centering $a=1.78$, $b=40.07$}  \\
        \hline
        \parbox[c]{0.12\linewidth}{\vspace{4pt}\centering \textit{Standalone \ac{DMC} arrival time} $[\textbf{\text{m}}]$\vspace{4pt}} 
        &\parbox[c]{0.08\linewidth}{\centering $\mathcal{G}(a,b)$ }
        &\textbf{--}
        &\parbox[c]{0.15\linewidth}{\centering $a=2.78$, $b=31.32$}
        &\parbox[c]{0.15\linewidth}{\centering $a=4.58$, $b=27.29$}  \\
        \hline
        \parbox[c]{0.12\linewidth}{\vspace{4pt}\centering \textit{MPC cluster angular centroids} $[\textbf{\text{rad}}]$\vspace{4pt}} 
        &\parbox[c]{0.08\linewidth}{\centering $\mathcal{BL}(w,\mu,\lambda)$ }
        &0
        &\parbox[c]{0.15\linewidth}{\centering $w=[0.46,0.54]$ \\ $\mu = [0,\pi]$ \\ $\lambda = [0.20,1.14]$}
        &\parbox[c]{0.13\linewidth}{\centering $U[0,2\pi)$}  \\
        \hline
        \parbox[c]{0.12\linewidth}{\vspace{4pt}\centering \textit{Standalone \ac{DMC} angular centroids} $[\textbf{\text{rad}}]$\vspace{4pt}} 
        &\parbox[c]{0.08\linewidth}{\centering $\mathcal{BL}(w,\mu,\lambda)$ }
        &0
        &\parbox[c]{0.15\linewidth}{\centering $w=[0.56,0.44]$ \\ $\mu = [0,\pi]$ \\ $\lambda = [0.29,0.56]$}
        &\parbox[c]{0.13\linewidth}{\centering $U[0,2\pi)$}  \\
        \hline
        \parbox[c]{0.12\linewidth}{\vspace{4pt}\centering \textit{Intra-cluster excess delay} $[\textbf{\text{m}}]$\vspace{4pt}} 
        &\parbox[c]{0.08\linewidth}{\vspace{4pt}\centering (\ref{eq:segmented_fit}) \vspace{4pt}}
        &\parbox[c]{0.15\linewidth}{\centering $\tau_{\rm BP}=0.30$ \\ $\lambda=[11.63,0.37]$}
        &\parbox[c]{0.15\linewidth}{\centering $\tau_{\rm BP}=0.37$ \\ $\lambda=[9.36,0.61]$}
        &\parbox[c]{0.15\linewidth}{\centering $\tau_{\rm BP}=0.29$ \\ $\lambda=[6.85,0.33]$}  \\
        \hline
        \parbox[c]{0.15\linewidth}{\vspace{4pt}\centering \textit{\ac{LoS} fading} $[\textbf{\text{rad}}]$\vspace{4pt}} 
        &\parbox[c]{0.1\linewidth}{\centering $\mathcal{R}(\nu,\sigma)$ }
        &\parbox[c]{0.15\linewidth}{\vspace{4pt}\centering $\nu = 0.91$, $\sigma=0.25$\vspace{4pt}}
        &\textbf{--}
        &\textbf{--}  \\
        \hline
        \parbox[c]{0.15\linewidth}{\vspace{4pt}\centering \textit{Intra-cluster angular offset} $[\textbf{\text{rad}}]$\vspace{4pt}} 
        &\parbox[c]{0.1\linewidth}{\centering $\mathcal{L}(\mu=0,\lambda)$ }
        &\parbox[c]{0.15\linewidth}{\vspace{4pt}\centering $\lambda = 0.14, \tau<\tau_{\rm BP}$ \\ $\lambda = 0.19, \tau>\tau_{\rm BP}$\vspace{4pt}}
        &\parbox[c]{0.15\linewidth}{\centering $\lambda = 0.10, \tau<\tau_{\rm BP}$ \\ $\lambda = 0.18, \tau>\tau_{\rm BP}$}
        &\parbox[c]{0.15\linewidth}{\centering $\lambda = 0.15, \tau<\tau_{\rm BP}$ \\ $\lambda = 0.24, \tau>\tau_{\rm BP}$}  \\
        \hline
        \parbox[c]{0.12\linewidth}{\vspace{4pt}\centering \textit{Intra-cluster power decay} $[\textbf{\text{dB}}]$\vspace{4pt}} 
        &\parbox[c]{0.08\linewidth}{\centering (\ref{eq:decay_model}) \\
        $(\zeta, \lambda_{\tau}, \lambda_{\phi})$}
        &\parbox[c]{0.22\linewidth}{\centering $(0.79, 24.28, 35.85), \tau<\tau_{\rm BP}$ \\ $(-30.99, 2.40, 1.76), \tau>\tau_{\rm BP}$}
        &\parbox[c]{0.22\linewidth}{\centering $(3.93, 9.31, 27.65), \tau<\tau_{\rm BP}$ \\ $(-15.26, 2.12, 3.60), \tau>\tau_{\rm BP}$}
        &\parbox[c]{0.22\linewidth}{\centering $(-14.25, 8.04, 7.10), \tau<\tau_{\rm BP}$ \\ $(-19.32, 0.06, 5.96), \tau>\tau_{\rm BP}$}  \\
        \hline
        \parbox[c]{0.12\linewidth}{\vspace{4pt}\centering \textit{Cluster mean power} $[\textbf{\text{dB}}]$\vspace{4pt}} 
        &\parbox[c]{0.08\linewidth}{\centering (\ref{eq:mean_power}) }
        &\textbf{--}
        &\parbox[c]{0.19\linewidth}{\centering $\alpha_{\rm cl}=-22.09$, $\beta_{\rm cl} = -1.35$}
        &\parbox[c]{0.19\linewidth}{\centering $\alpha_{\rm cl}=-28.89$, $\beta_{\rm cl} = -2.17$}  \\
        \hline
        \parbox[c]{0.12\linewidth}{\vspace{4pt}\centering \textit{Cluster power std} $[\textbf{\text{dB}}]$\vspace{4pt}} 
        &\parbox[c]{0.08\linewidth}{\centering (\ref{eq:power_std}) }
        &\textbf{--}
        &\parbox[c]{0.19\linewidth}{\centering $\rho_{\rm cl}=7.89$, $\eta_{\rm cl} = -0.24$}
        &\parbox[c]{0.19\linewidth}{\centering $\rho_{\rm cl}=6.97$, $\eta_{\rm cl} = 0.13$}  \\
        \hline
        \parbox[c]{0.12\linewidth}{\vspace{4pt}\centering \textit{Standalone \ac{DMC} mean power} $[\textbf{\text{dB}}]$\vspace{4pt}} 
        &\parbox[c]{0.08\linewidth}{\centering (\ref{eq:mean_power}) }
        &\textbf{--}
        &\parbox[c]{0.19\linewidth}{\centering $\alpha_{\rm cl}=-31.57$, $\beta_{\rm cl} = -1.43$}
        &\parbox[c]{0.19\linewidth}{\centering $\alpha_{\rm cl}=-41.27$, $\beta_{\rm cl} = -1.03$}  \\
        \hline
        \parbox[c]{0.12\linewidth}{\vspace{4pt}\centering \textit{Standalone \ac{DMC} power std} $[\textbf{\text{dB}}]$\vspace{4pt}} 
        &\parbox[c]{0.08\linewidth}{\centering (\ref{eq:power_std}) }
        &\textbf{--}
        &\parbox[c]{0.19\linewidth}{\centering $\rho_{\rm cl}=6.57$, $\eta_{\rm cl} = 0.44$}
        &\parbox[c]{0.19\linewidth}{\centering $\rho_{\rm cl}=7.39$, $\eta_{\rm cl} = 0.21$}  \\
        \hline
        \parbox[c]{0.12\linewidth}{\vspace{4pt}\centering \textit{\ac{SDR}} $[\textbf{\text{dB}}]$\vspace{4pt}} 
        &\parbox[c]{0.08\linewidth}{\centering $\mathcal{N}(\mu,\sigma)$ }
        &\parbox[c]{0.15\linewidth}{\centering $\mu=9.90$, $\sigma = 2.45$}
        &\parbox[c]{0.15\linewidth}{\centering $\mu=2.19$, $\sigma = 3.56$}
        &\parbox[c]{0.15\linewidth}{\centering $\mu=-1.09$, $\sigma = 5.42$}  \\
        \hline
    \end{tabular}
    \label{tab:summary}
    \vspace{-1em}
\end{table*}

\begin{table}[tb]
\centering
\caption{Standalone \ac{DMC} Spread Modeling}
\label{tab:sDMC_spread}
\begin{tabular}{llll}
\toprule
\textbf{Statistics} & \textbf{Modeling} &{\centering \ac{LoS}} &{\centering \ac{NLoS}} \\
\midrule
Delay &$\mathcal{G}(a,b)$ &$a=5.03$, $b=0.22$ &$a=8.31$, $b=0.16$ \\
\midrule
Angular &$\mathcal{G}(a,b)$ &$a=5.15$, $b=0.03$ &$a=5.87$, $b=0.05$ \\
\bottomrule
\end{tabular}
\end{table}

\begin{table}[tb]
\centering
\caption{Associated \ac{DMC} Spread Modeling}
\label{tab:aDMC_spread}
\begin{tabular}{llll}
\toprule
\textbf{Statistics} & \textbf{Modeling} &{\centering direct/\ac{LoS}} &{\centering \ac{NLoS}} \\
\midrule
Delay &$\mathcal{G}(a,b)$ &$a=4.87$, $b=0.27$ &$a=7.66$, $b=0.19$ \\
\midrule
Angular &$\mathcal{G}(a,b)$ &$a=4.91$, $b=0.10$ &$a=5.86$, $b=0.09$ \\
\bottomrule
\end{tabular}
\vspace{-12pt}
\end{table}

\vspace{-12pt}
\section{Synthetic Channel Generation }

The extracted statistics form the basis of a channel model that can generate double-directional impulse responses that are consistent with our measurement campaign, but also enables the generation of many more channel realizations than the number of location pairs, or with other system parameters such as beamwidth. The following describes the flow of the impulse response generation; a MATLAB implementation is made publicly available as part of this publication. 

The user input of the algorithm are the geometrical distance between Tx and Rx (must be 1--100 m to be consistent with the measurement range), the propagation condition (LoS or NLoS scenario), and the bandwidth, antenna pattern, and antenna pointing directions desired by users.\footnote{Additional oversampling and filtering can be applied after the generation of the double-directional impulse response.}
The next step is the generation of the number of MPC clusters and DMC clusters drawn from Gamma distributions with the parameters in Sec. II, and generating for MPC-associated DMC clusters, and standalone DMC clusters, the arrival time and DoA and DoD of the cluster centroids, according to their respective distributions. Note that the arrival times and centroid angles of the associated DMC clusters coincide with those of their corresponding MPC clusters. To avoid overlap between clusters, if a generated standalone DMC falls within a predefined local region of an existing MPC cluster center across all three parametric domains, it is rejected and regenerated. This ensures that standalone DMCs remain separated from other clusters, consistent with the measurement observations.

In the next step, we generate the MPCs. In particular, for each cluster, we first randomly draw the number of MPCs from the Gamma distribution characterizing the ``number of MPCs per cluster''. For each of the selected MPCs, we then determine the offset in the delay domain, according to (\ref{eq:segmented_fit}), and the offset in DoA and DoD from a Laplacian distribution (the width of that distribution depends on whether the delay is larger or smaller than the breakpoint). For each LoS or NLoS cluster, the 
power is determined according to the intra-cluster power decay law plus random variations around it. 
We then first compute the desired total cluster power (MPC+DMC) from a realization of (11+12), and from this the pure cluster MPC power by determining the MPC/DMC ratio as per Sec. IV.M). We finally renormalize all MPCs so that their sum power equals this MPC cluster power value. 

For the direct cluster, the procedure is slightly different, since extraction of cluster power statistics is difficult due to the small number of measured direct clusters. We first generate the amplitude of the direct MPC, $a_0$ as a realization of a Rician variable with mean 0.91 (note that at this stage, all power values are normalized to FSPL at TX-RX distance). We then generate $M$ MPCs (this number follows as realization of the process in Sec. IV.B), at randomly chosen delay- and angle offsets (according to Sec. IV.E/F), and generate their powers according to (9), to provide normalized path amplitudes $\tilde{a_i}, i=1...M$. We then determine $\tilde{a_0}$ from the condition $\sum_{i=0}^{M}|\tilde{a_i}|^2=1$.\footnote{If $\sum_{i=0}^{M}|\tilde{a_i}|^2>1$, the realizations of the MPCs are discarded and another realization is generated.} We then re-normalize all MPC powers by multiplying them with $\sqrt{FSPL}\cdot a_0/\tilde{a}_0$, from which we then obtain the total pure MPC cluster power. The DMC power of the direct cluster is then computed via a realization of the SDR distribution. 



The next step is the determination of the cluster powers. The nominal direct-component power, which is used as reference value for subsequent computations, can be computed deterministically from the Tx-Rx distance. Then, for each LoS and NLoS cluster, determine the excess attenuation as realization of (\ref{eq:mean_power}) and (\ref{eq:power_std}), and use it to normalize the power of each MPC and DMC tap. For each DMC tap (in both associated and standalone clusters), a realization of the amplitude is then selected according to the mean power value that follows from the total cluster power and the power decay as a function of delay and angle offset, as parameterized in Table II. Finally, threshold the DDIR with the desired dynamic range (no more than 35 dB). This DDIR can then be combined with arbitrary beam/antenna patterns to obtain transfer functions or transfer function matrices to be used in further system simulations.


\vspace{-10pt}
\section{Model Validation and Verification }

The accuracy of the statistical fits of the {\em extracted} parameters used to build the model were given in Sec. IV. Next, we verify whether the model can also accurately match the statistics of the raw measurements for {\em derived} channel characteristics, i.e., those that are not directly fitted.\footnote{A different problem is the verification that the model including its parameterization is capable of describing different environments. This type of verification needs another large set of measurements. Given the difficulties of performing such campaigns in this frequency range (our measurements took a year), such an investigation is outside the scope of this paper}. Significant deviations could occur due to (i) neglecting important correlations between extracted parameters, and (ii) error accumulation from the different component statistics. However,  the validation results reported in the next section indicate that the retained statistics are sufficient to reproduce the dominant channel characteristics. The following subsections test statistics of some key derived parameters.

\begin{figure*}[tb]
    \centering
    \setcounter{subfigure}{0}
    \begin{subfigure}[t]{0.32\textwidth}
        \centering
        \includegraphics[width=\linewidth]{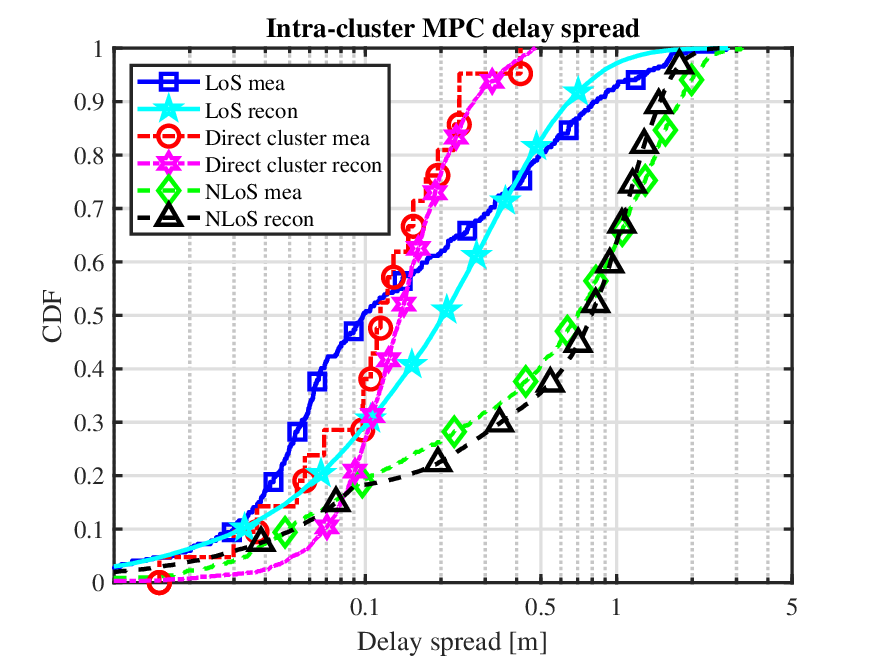}
        \caption{Intra-cluster \ac{MPC} delay spread comparison.}
        \label{fig:intra_DS_MPC_comp}
    \end{subfigure}
    \hfill
    \setcounter{subfigure}{2}
    \begin{subfigure}[t]{0.32\textwidth}
        \centering
        \includegraphics[width=\linewidth]{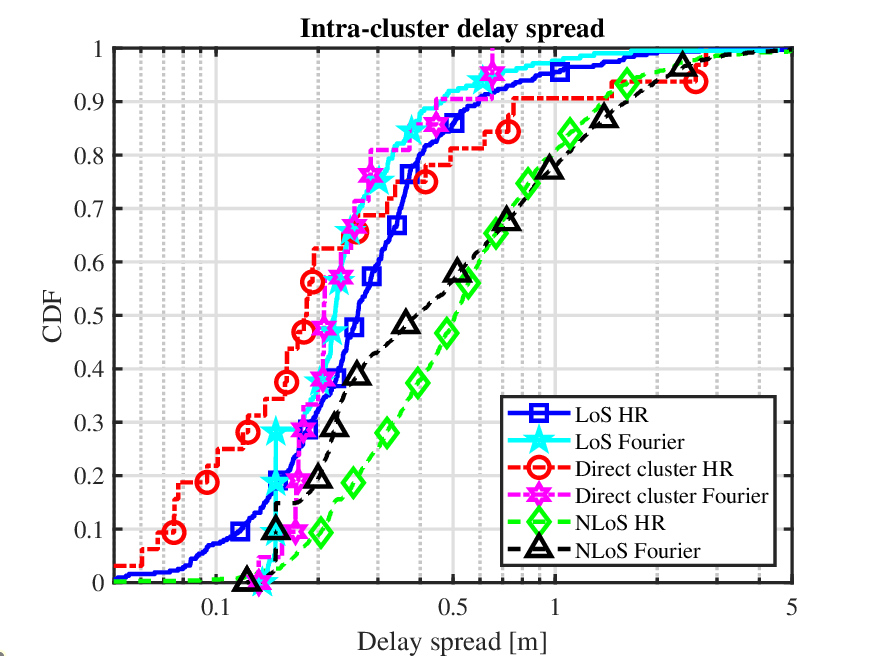}
        \caption{Intra-cluster delay spread comparison.}
        \label{fig:intra_cluster_DS_comp}
    \end{subfigure}
    \hfill
    \setcounter{subfigure}{4}
    \begin{subfigure}[t]{0.32\textwidth}
        \centering
        \includegraphics[width=\linewidth]{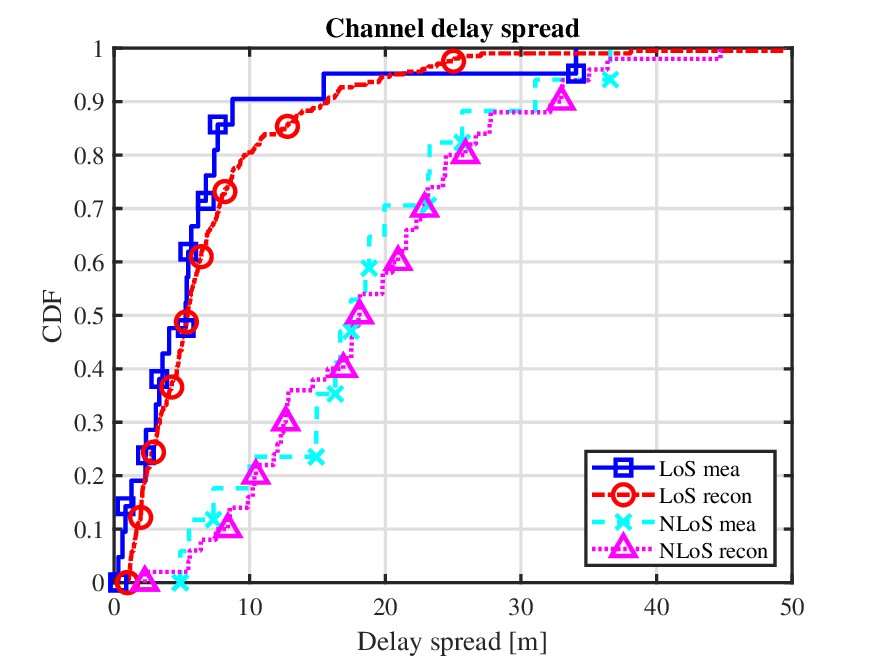}
        \caption{Channel delay spread comparison.}
        \label{fig:channel_DS}
    \end{subfigure}
    \\
    \centering
    \setcounter{subfigure}{1}
    \begin{subfigure}[t]{0.32\textwidth}
        \centering
        \includegraphics[width=\linewidth]{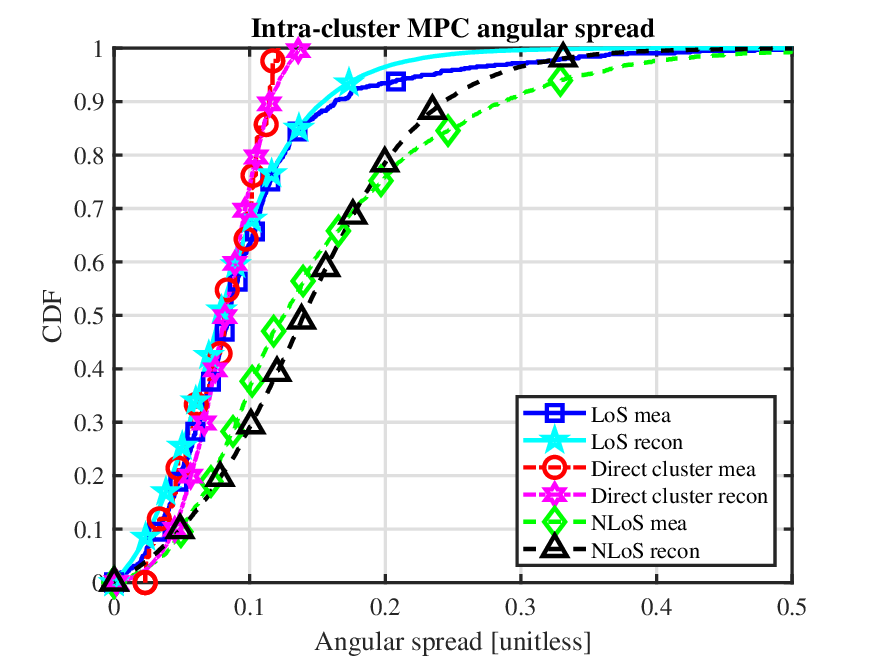}
        \caption{Intra-cluster \ac{MPC} angular spread comparison.}
        \label{fig:intra_AS_MPC_comp}
    \end{subfigure}
    \hfill
    \setcounter{subfigure}{3}
    \begin{subfigure}[t]{0.32\textwidth}
        \centering
        \includegraphics[width=\linewidth]{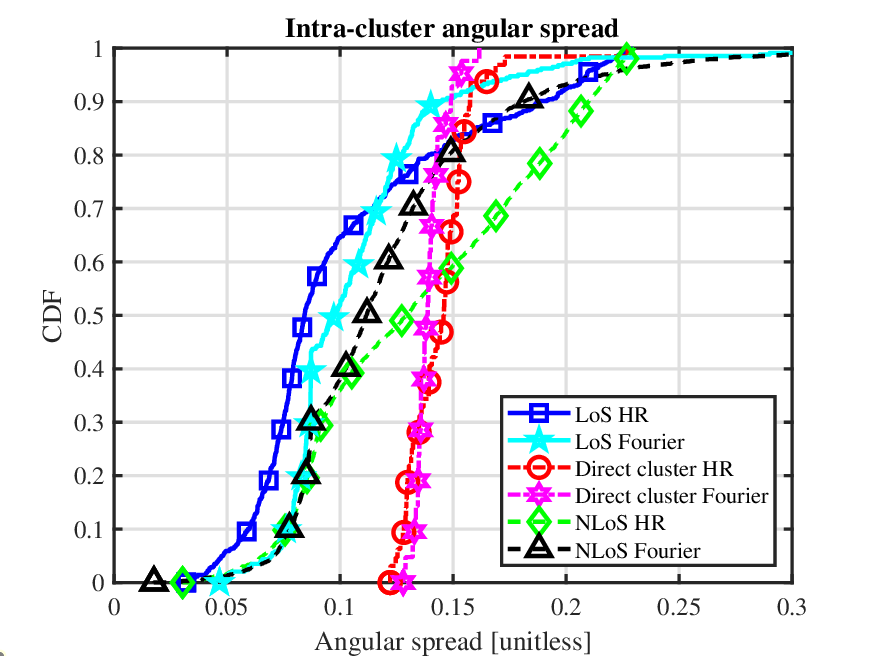}
        \caption{Intra-cluster angular spread comparison.}
        \label{fig:intra_cluster_AS_comp}
    \end{subfigure}
    \hfill
    \setcounter{subfigure}{5}
    \begin{subfigure}[t]{0.32\textwidth}
        \centering
        \includegraphics[width=\linewidth]{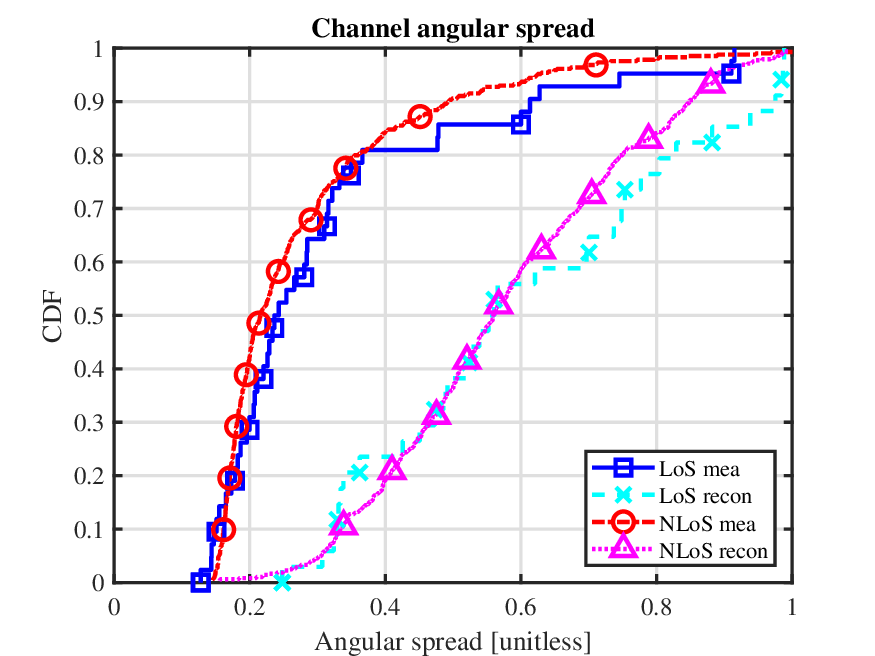}
        \caption{Channel angular spread comparison.}
        \label{fig:channel_AS}
    \end{subfigure}
    \caption{Validation of the proposed synthetic channel generation framework by comparing key intra-cluster \ac{MPC} and channel statistics between measured and synthetic channels: (a,c,e) delay spread, and (b,d,f) angular spread.}
    \label{fig:statistics_comp}
    \vspace{-1em}
\end{figure*}

\vspace{-16pt}
\subsection{Intra-Cluster Power Concentration of Dominant MPCs}

\begin{figure}[tb]
\centering
\includegraphics[width=\linewidth]{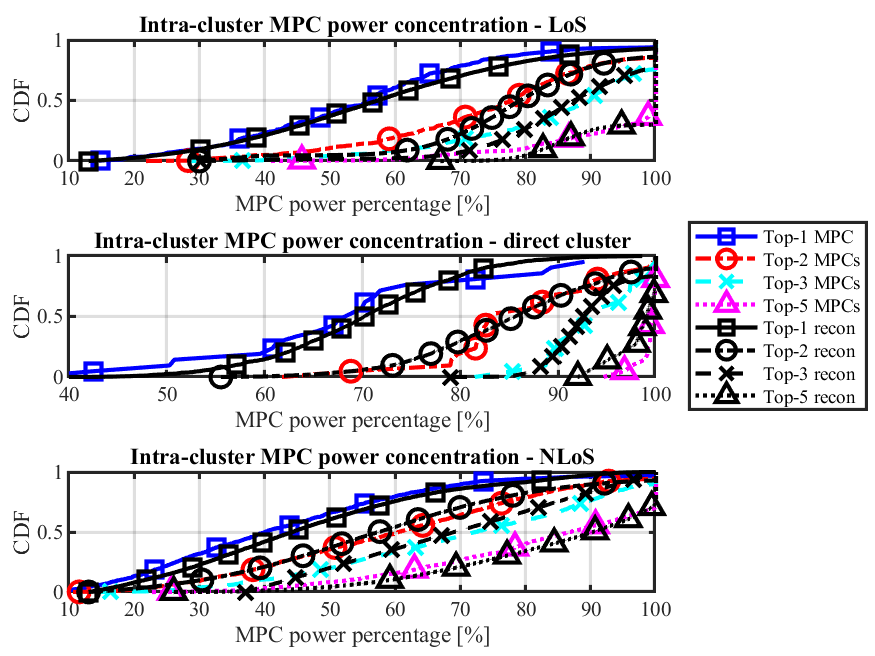}
\caption{The power contained in the strongest 1,2,3, and 5 MPCs under \ac{LoS}, \ac{LoS}-cluster, and \ac{NLoS} conditions.}
\label{fig:MPC_power_concentration}
\vspace{-1em}
\end{figure}

The power concentration of MPCs in a cluster describes how many MPCs it takes to explain a certain percentage of energy in a cluster (this only refers to discrete MPCs, since DMC per-tap power depends on the chosen sample spacing in delay/angle). 
We sort the MPCs according to their power $P_{(1)} \geq \cdots \geq P_{(k)} \geq \cdots$, and then compute $\sum_{k=1}^{K}P_{(k)}/\sum_{k=1}^{N_c^{\rm MPC}}P_{(k)}$.
Fig.~\ref{fig:MPC_power_concentration} shows the CDF of the power contained in the strongest $K=1,2,3, 5$ MPCs in direct, LoS, and NLoS clusters. Not surprisingly, for the direct cluster, the smallest number of MPCs is required to account for a given energy percentage, with LoS clusters requiring more, and NLoS clusters the most. Still, in about half of the cases, just one (direct and LoS) or two (NLos) MPCs are sufficient to explain half the energy. For the direct cluster, 5 MPCs can explain 90\% of the power.

Another way to quantify the power concentration is the intra-cluster decay of the MPC power. We again sort the MPCs in a cluster according to their power, and then analyze the decay of $P_{(k)}$ with $k$. From analysis of the RiMAX-extracted values, we find that an exponential decay well describes this behavior, with the decay constant varying from cluster to cluster. We thus normalize the received power to the total cluster power, and perform a least-square fit of the sorted values in the dB domain. The extracted slope (decay constant) then characterizes the power concentration, with larger values representing a stronger concentration of the power.

\begin{figure}[tb]
\centering
\includegraphics[width=0.75\linewidth]{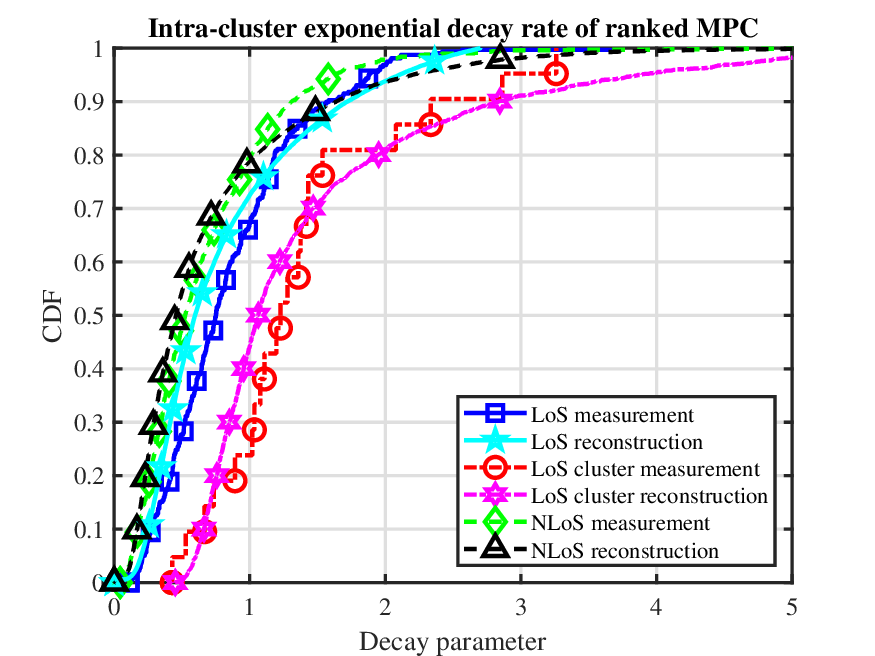}
\caption{Validation of the proposed synthetic channel generation framework by comparing intra-cluster decay parameter of ranked \ac{MPC} between measured and synthetic channels.}
\label{fig:intra_MPC_rank_decay}
\vspace{-1em}
\end{figure}

Fig.~\ref{fig:intra_MPC_rank_decay} shows the CDF of the decay parameter for direct, LoS, and NLoS clusters. We can see good agreement between the raw measurements and the values obtained from synthesized impulse responses, with CvM values of 0.125, 0.007, 0.043, respectively.  Decay constants at the median level is larger than 1 for the direct cluster (i.e., the second-strongest component is only 30\% of the power of the strongest component), and 0.5 for the NLoS case. 

These numbers are in stark contrast to the modeling of MPC powers in the 3GPP model, where 20 equi-powered MPCs are assumed in each cluster. Since 3GPP recently introduced the option of reducing the number of MPCs per cluster for future models (e.g., new frequency range), the results of our investigation can be particularly relevant.

\vspace{-10pt}
\subsection{Intra-Cluster Statistics}
\vspace{-4pt}

Intra-cluster delay spread and angular spread are used as ``primary" statistics (i.e., directly fitted) in many papers. However, in our work we {\em separately} model MPC density and MPC power. We thus need to verify whether this split (and the assumption that the distribution for these two quantities are independent) properly reproduces the conventional intra-cluster spread distributions. We thus compare the CDFs of the spreads as computed from the raw extracted MPCs with those obtained from the MPCs generated with the computer program implementing our model. The results are shown in Figs.~\ref{fig:intra_DS_MPC_comp} and \ref{fig:intra_AS_MPC_comp}, for rms delay and angular spreads, respectively. 
They visually demonstrate good agreement across all scenarios. This is quantitatively supported by the \ac{CvM} distances of 0.006, 0.177, and 0.248 for the delay spread in the Direct, LoS, and NLoS cases, respectively, and 0.135, 0.010, and 0.084 for the corresponding angular spread.

We also compare intra-cluster delay and angular spreads from HRPE vs Fourier analysis results from \cite{cheng2025stochastic}. Fig.~\ref{fig:intra_cluster_DS_comp} and \ref{fig:intra_cluster_AS_comp} show this comparison of different modeling frameworks. We see somewhat larger deviations. This is caused mainly by three effects: (i) the combination of MPC and DMC, each of which have errors with respect to the raw extracted data; (ii) the different clustering, as the clustering methodology is necessarily different when distinguishing MPC and DMC clusters, and (iii) the finite resolution of the Fourier method which makes very small spread values less likely. Still, we notice qualitative agreement, in particular in terms of the median and slope of the curves.  

\vspace{-10pt}
\subsection{Channel Statistics}

The final check of our model is the \ac{CDF} of the {\em overall} \ac{RMS} delay spread and angular spread comparing model-generated results to the \acp{CDF} computed directly from the measurement data.
As these quantities incorporate both inter-and intra-cluster statistics, it is a good indicator of any possible significant omissions in the modeling process. Fig.~\ref{fig:channel_DS} and \ref{fig:channel_AS} show the results. Again, we see visually good agreement, which is quantitatively supported by the \ac{CvM} distances of 0.009 and 0.008 for the delay spread in \ac{LoS} and \ac{NLoS} scenarios, respectively, and 0.110 and 0.106 for the angular spread in LoS and NLoS, respectively.

Since the rms delay and angular spread are not influenced by the scaling of the total power in the impulse response, Table \ref{tab:linear_model} compares the pathloss and shadowing variance obtained from our model versus that obtained from directly fitting the measurements in \cite{abbasi2022thz}. Note that our model does {\em not} do a renormalization of the total power according to bulk statistics, but rather generates cluster powers according to the description in Sec. V, and the sum of the cluster powers provides the generated total power. The obtained values for both omni-directional and max-dir (best orientations of the horn antennas at both link ends) pathloss are in good agreement, and in particular fall within the confidence intervals. This holds both for the parameters of an $\alpha-\beta$ pathloss model and a close-in pathloss model.

\begin{table*}[tb]
\vspace{10pt}
    \centering
    {
    \caption{The comparison of path loss parameters}
    \makebox[\linewidth][c]{\resizebox{0.8\linewidth}{!}{
    \begin{tabular}{|c|c|c|c|c|c|c|c|c|c|c|c|c|}
        \hline
        \rule[-1.5ex]{0pt}{4ex} \textbf{Parameters} &$\alpha$ in \cite{abbasi2022thz} &$\alpha$ &$\alpha_{\rm 95\%~CI}$ &$\beta$ in \cite{abbasi2022thz} &$\beta$ &$\beta_{\rm 95\%~CI}$ &$\mu$ in \cite{abbasi2022thz} &$\mu$ &$\mu_{\rm 95\%~CI}$ &$\sigma$ in \cite{abbasi2022thz} &$\sigma$ &$\sigma_{\rm 95\%~CI}$ \\
        \hline
        \hline
        \rule[-1.5ex]{0pt}{4ex} $\text{PL}_{\rm omni}^{\rm LoS}$ &76.86 &75.98 &[74.00,77.96] &1.77 &1.81 &[1.69,1.93] &0 &-1.12 &[-3.58,1.38] &1.28 &1.68 &[1.26,2.34]\\
        \hline
        \rule[-1.5ex]{0pt}{4ex} $\text{PL}_{\rm max-dir}^{\rm LoS}$ &76.68 &76.14 &[72.56,79.72] &1.84 &1.87 &[1.53,2.21] &0 &-2.56 &[-5.98,2.88] &1.63 &2.50 &[1.79,3.33]\\
        \hline
        \rule[-1.5ex]{0pt}{4ex} $\text{PL}_{\rm omni}^{\rm NLoS}$ &95.84 &95.82 &[88.33,103.30] &1.57 &1.50 &[1.01,2.17] &0.33 &-2.39 &[-5.08,0.64] &5.07 &8.05 &[4.84,11.26] \\
        \hline
        \rule[-1.5ex]{0pt}{4ex} $\text{PL}_{\rm max-dir}^{\rm NLoS}$ &99.69 &100.42 &[90.87,110.68] &1.54 &1.59 &[0.84,2.37] &0.38 &-1.99 &[-4.58,0.94] &7.03 &8.77 &[6.93,11.00] \\
        \hline
    \end{tabular}
    }}
    
    \par\vspace{2mm}

    \makebox[\linewidth][c]{\resizebox{0.8\linewidth}{!}{
    \begin{tabular}{|c|c|c|c|c|c|c|c|c|}
        \hline
        \rule[-1.5ex]{0pt}{4ex} \textbf{Parameters} &$n^{\rm mea}$ &$n_{\rm 95\%~CI}^{\rm mea}$ &CIM $\sigma^{\rm mea}$ &CIM $\sigma_{\rm 95\%~CI}^{\rm mea}$ &$n^{\rm fit}$ &$n_{\rm 95\%~CI}^{\rm fit}$ &CIM $\sigma^{\rm fit}$ &CIM $\sigma_{\rm 95\%~CI}^{\rm fit}$\\
        \hline
        \hline
        \rule[-1.5ex]{0pt}{4ex} $\text{PL}_{\rm omni}^{\rm LoS}$ &1.83 &[1.78,1.87] &1.37 &[1.05,1.96] &1.86 &[1.83,1.90] &2.00 &[1.67,2.48] \\
        \hline
        \rule[-1.5ex]{0pt}{4ex} $\text{PL}_{\rm max-dir}^{\rm LoS}$ &1.90 &[1.84,1.96] &1.83 &[1.41,2.62] &1.88 &[1.85,1.92] &2.04 &[1.70,2.53] \\
        \hline
        \rule[-1.5ex]{0pt}{4ex} $\text{PL}_{\rm omni}^{\rm NLoS}$ &2.86 &[2.64,3.08] &7.38 &[5.54,11.07] &2.92 &[2.78,3.07] &8.05 &[6.74,10.00] \\
        \hline
        \rule[-1.5ex]{0pt}{4ex} $\text{PL}_{\rm max-dir}^{\rm NLoS}$ &3.05 &[2.77,3.33] &9.33 &[7.01,14.00] &3.18 &[3.02,3.35] &9.29 &[7.78,11.55] \\
        \hline
    \end{tabular}
    \label{tab:linear_model}
    }}
    }
\end{table*}

\section{Conclusion}

This paper performed a HRPE analysis of double-directional propagation channels in the sub-THz frequency band for a D2D deployment scenario. Using a generalization of the RiMAX algorithm, we extracted both specular and diffuse multipath, clustered the contributions, and provided complete intra- and inter-cluster statistics. This provides, to our knowledge, the first high-resolution model for outdoor sub-THz channels; we also make publicly available a MATLAB program that generates channel realizations according to this model. 

Besides the new model parameterization, we also introduce several conceptual novelties, including the splitting of DMC clusters into associated and free-standing clusters, the separate parameterization of MPC density and power from measurements, and the separate parameterization of the direct cluster and the remaining clusters in LoS scenarios. We find that the power-sorted MPCs show a significant decay, indicating that the assumption of multiple equi-powered MPCs (as used, e.g., in the 3GPP  channel model) does not hold in the measured environment. Extensive verifications demonstrate that quantities that are not directly modeled (e.g., overall rms delay and angular spread and pathloss slopes) still are reproduced correctly by our model. For future work, we will apply the modeling framework to other measurement campaigns to assess how dependent the extracted parameters are on the specific environment. 

\bibliographystyle{ieeetr}
\bibliography{ref}

\end{document}